\documentclass[a4paper,fleqn,usenatbib]{mnras}

\usepackage{newtxtext,newtxmath}

\usepackage{lscape}
\usepackage{hyperref}
\usepackage[flushleft]{threeparttable}

\usepackage{graphicx}	
\usepackage{amsmath}	
\usepackage{amssymb}	

\newcommand\Minit{M$_0$}
\newcommand\Mzams{M$_0$(env)}
\newcommand\Msin{M$_0$(env)$_{\rm{sin}}$}
\newcommand\Mbin{M$_0$(env)$_{\rm{bin}}$}
\newcommand\Mcor{M$_0$(env)$_{\rm{corr}}$}
\newcommand\Mlit{M$_0$(direct)}
\newcommand{\ha}{H$\alpha$}
\newcommand{\hb}{H$\beta$}
\newcommand{\hii}{\ion{H}{ii}}

\title[Muse observations of the nearby environments of Type II SNe]{The 50$-$100~pc scale parent stellar populations of type II supernovae and limitations of single star evolution models}

\author[P. Schady et al.]{P.~Schady$^{1,2}$\thanks{E-mail: p.schady@bath.ac.uk (PS)}
, J.J.~Eldridge$^3$, J.~Anderson$^4$, T.-W.~Chen$^2$, L.~Galbany$^{5,6}$, \newauthor H~Kuncarayakti$^{7,8}$ and L.~Xiao$^{9,10}$
\\
$^1$ Department of Physics, University of Bath, Claverton Down, Bath, BA2 7AY, UK\\
$^2$ Max-Planck-Institut f{\"u}r Extraterrestrische Physik, Giessenbachstra\ss e, 85748, Garching, Germany\\
$^3$ Department of Physics, Private Bag 92019, University of Auckland, Auckland 1010, New Zealand\\
$^4$ European Southern Observatory, Alonso de C{\'o}rdova 3107, Vitacura, Casilla 19001, Santiago, Chile\\
$^5$ Departamento de F\'isica Te\'orica y del Cosmos, Universidad de Granada, E-18071 Granada, Spain \\
$^6$ PITT PACC, Department of Physics and Astronomy, University of Pittsburgh, Pittsburgh, PA 15260, USA\\
$^7$ Finnish Centre for Astronomy with ESO (FINCA), University of Turku, V{\"a}is{\"a}l{\"a}ntie 20, 21500 Piikki{\"o}, Finland\\
$^8$ Tuorla Observatory, Department of Physics and Astronomy, University of Turku, V{\"a}is{\"a}l{\"a}ntie 20, 21500 Piikki{\"o}, Finland\\
$^9$ CAS Key Laboratory for Research in Galaxies and Cosmology, Department of Astronomy, University of Science and Technology of China, Hefei, 230026, China\\
$^{10}$ School of Astronomy and Space Sciences, University of Science and Technology of China, Hefei, 230026, China
}

\date{Accepted 2019 October 3. Received 2019 September 20; in original form 2019 July 22}

\pubyear{2019}

\begin{document}
\label{firstpage}
\pagerange{\pageref{firstpage}--\pageref{lastpage}}
\maketitle

\begin{abstract}
There is observational evidence of a dearth in core-collapse supernova (ccSN) explosions from stars with zero age main sequence (ZAMS) mass \Minit$\approx17-30 \mathrm{M}_\odot$, referred to as the `red supergiant problem'. However, simulations now predict that above 20~M$_\odot$ we should indeed only expect stars within certain pockets of \Minit\ to produce a visible SN explosion. Validating these predictions requires large numbers of ccSNe of different types with measured \Minit, which is challenging. In this paper we explore the reliability of using host galaxy emission lines and the \ha\ equivalent width to constrain the age, and thus the \Minit\ of ccSNe progenitors. We use Binary Population and Spectral Synthesis models to infer a stellar population age from MUSE observations of the ionised gas properties and \ha\ EW at the location of eleven ccSNe with reliable \Minit\ measurements. Comparing our results to published \Minit\ values, we find that models that do not consider binary systems yield stellar ages that are systematically too young (thus \Minit\ too large), whereas accounting for binary system interactions typically overpredict the stellar age (thus underpredict \Minit). Taking into account the effects of photon leakage bring our \Minit\ estimates in much closer agreement with expectations. These results highlight the need for careful modelling of diffuse environments, such as are present in the vicinity of type II SNe, before ionised emission line spectra can be used as reliable tracers of progenitor stellar age.
\end{abstract}




\section{Introduction}
Stars with \Minit $\ga 8~\mathrm{M}_\odot$ end their lives in the collapse of their dense stellar cores, typically resulting in a cataclysmic explosion. The precise end product at the time of explosion is a sensitive function of primarily the progenitor zero age main sequence (ZAMS) mass, but also metallicity, giving rise to a range of SN explosions. However, whereas models predicted that all stars with \Minit\ up to 40~M$_\odot$ should end their lives as a SN \citep[and at greater masses in metal-rich environments;][]{hfw+03}, observations imply that there is a cap in the progenitor \Minit\ at $\sim 20$~M$_\odot$ \citep[][although see \citet{db18} for a possible cap as high as \Minit$<33~\mathrm{M}_\odot$, 95\% confident]{sec+09,jfm+12,jsf+14,jes+15,sma15,spw+17}. More recently, detailed simulations have shown that above $\sim 20$~M$_\odot$, it becomes much harder for the energy released during core-collapse to escape the stellar envelope and produce a SN, resulting in `islands of explodability' \citep[see][for review]{jan12}. The models imply that, instead, these `failed' SN collapse directly to form a black hole \citep{jan12,lw13,pir13,sew+16,swh+18}, possibly linked to a change in the nature of carbon burning in very massive stars. Surveys have been developed to try and find failed SN candidates by searching for the disappearance of evolved stars in nearby galaxies \citep{kbk+08,rfg+15}. However, identification and verification of candidates is challenging \citep{gks15}, with thus far just two promising candidates having been found over the past five years \citep{gks15,rfg+15,akg+17}. In light of these theoretical developments on the fate of massive stars, far greater samples of SNe with \Minit\ estimates are required to try and find these islands of explodability from observations.

The most accurate observational estimates of \Minit\ come from direct detections of the progenitor star from pre-explosion, high resolution images \citep[predominantly taken with the Hubble Space Telescope; e.g.][]{sec+09,fee+11,vlc+11,sma15, van17}. From the observed progenitor colours, it is then possible to infer the initial mass of the star from its derived luminosity, using stellar evolution models. Relatively reliable mass estimates can also be obtained from modelling the SN spectrum once it has reached the nebular phase. At this point, the ejecta is optically thin and the material at the core of the star can be seen, enabling the He core mass to be well constrained, which can be used to estimate the progenitor initial mass \citep{jfm+12,dhw+13,jsf+14,jes+15,fst+16,spw+17,adg+18}. Nevertheless, these more direct techniques rely on high quality data, requiring high spatial resolution to directly detect the progenitor star, or very sensitive, high signal-to-noise nebular spectra. These conditions therefore limit the use of these more direct techniques to the nearest, and brightest SNe.

An alternative technique that does not require observations of the SN or progenitor star is to age-date the young stars in the vicinity of the SN explosion, which are assumed to be the progenitor star's parent population. This type of analysis requires high spatial resolution data, generally taken with {\em HST}, using colour-magnitude diagrams to infer the ages of a few dozen young stars in the vicinity of recent SNe \citep{mjw+11,mau17,wpm+14,whm+18, alb+19} or SN remnants \citep[SNRs;][]{jwm+14,dmr+18}. The results have shown reduced evidence for an upper mass cut-off in the mass function of SNe II progenitors, although the uncertainties on \Minit\ are relatively large. Furthermore, the need for high spatial resolution, sensitive data, limits the use of this technique to the nearest SNe (typically type II), as is the case with the direct detection of the progenitor star. The sample sizes thus remain small, and with surprisingly little overlap with those SNe with initial masses obtained from progenitor or SN nebular spectral modelling \citep[around eight; e.g.][]{whm+18}.

In a similar vein to the age-dating method, information on the age of the underlying stellar population can also be gauged from modelling the gas emission line properties at the vicinity of the SN, which is ionised by nearby massive stars. Photoionisation from young stars produces strong \ha\ and \hb\ line emission, whereas the continuum emission is dominated by stellar light from older stars. The equivalent width of these hydrogen Balmer lines can thus be used to trace the age of the underlying, dominant stellar population \citep{lgs+11,kda+13a,kda+13b,gsm+14,gar+16,kag+18}. Larger EWs are indicative of younger stellar populations, and the EW of the \ha\ line, for example, is sensitive to stellar ages up to around 30~Myr. An advantage of this technique is that it is not necessary to resolve individual stars, making it applicable out to further distances. The distance limit is instead set by the need to resolve star forming regions dominated by a single stellar population, which observations imply is $\sim 100$~pc \citep{gp97,gdw+09,lkb+11,mjw+11}.

How the \ha\ EW relates to stellar age is dependent on the underlying model assumptions made, firstly on the properties of the stellar populations themselves (e.g. metallicity, binary fraction, initial mass function, rotation), and then on the conditions of the surrounding gas (e.g. electron density, ionisation parameter, UV background). With the same spectral data necessary to measure the \ha\ EW it is possible to also derive certain characteristic properties of the surrounding \hii\ region, such as the gas-phase metallicity, dust extinction, and electron density \citep[e.g.][]{xge+19}, which can be used to limit the sample size of synthetic spectra that are fitted to the data, thus reducing the number of degenerate parameters. Despite these observational constraints, there nevertheless remain uncertainties in how accurate the approximations inherent to stellar evolution models and in the conditions of the surrounding environment are. For example, most synthetic spectra assume spherical \hii\ region geometries, and that all ionising stellar light interacts with gas before escaping the \hii\ region, and the effects of dust are often neglected. In addition, post-main sequence stellar evolution, binary interactions and rotation can all significantly alter the results from population synthesis models \citep[e.g.][]{le12}. Verifying the model assumptions requires sensitive observations that resolve single stellar populations as best as possible, and which can break degeneracies between model parameters \citep[see][for a review on the limitations of determining physical properties from synthetic models]{con13}.

Given the implied potential that using \ha\ EW to trace progenitor age has on significantly increasing samples of SNe with measured \Minit, in this paper we wish to test some of the inherent assumptions involved when using the \ha\ EW as a tracer of the SN progenitor age. To do this we select a sample of SN with \Minit\ estimates reported in the literature, either from progenitor detections or SN nebular spectral modelling, and compare these published masses with the \Minit\ derived from the \ha\ EW. We use the VLT Multi-Unit Spectroscopic Explorer \citep[MUSE; ][]{baa+10} to study the environmental conditions in the vicinity of a sample of eleven SNe. MUSE provides imaging data over a $1\arcmin \times 1\arcmin$ field of view, with spectral information at every 0.2 arcsec sized pixel, allowing the spectral properties at the SN position and in its surrounding environment to be studied in detail. In section~\ref{sec:sample} we describe our sample selection criteria, and our MUSE data reduction is summarised in section~\ref{sec:analysis}. Our analysis and results are given in section~\ref{sec:results}, where we calculate various environmental properties at the SN position, and briefly describe the stellar population models used to infer \Minit\ from the measured \ha\ EW. In this section we also compare our \Minit\ estimates with initial masses reported in the literature from progenitor detections and nebular spectral modelling. Finally, in section~\ref{sec:disc} we discuss the reliability of using \ha\ EW and other emission lines to trace the age of stellar populations, and the limitations of stellar population models that only consider single stellar evolution. We finish with a short summary and conclusions in section~\ref{sec:concl}.

\section{Sample Selection}
\label{sec:sample}
The prime focus of this paper is to test the reliability of ccSN progenitor age and initial mass estimates from \ha\ EW measurements, and to explore possible modifications in the model assumptions that may improve the accuracy of this method. In order to carry out a cross check on \Minit, our first selection criteria is that the ccSN in our sample must have an \Minit\ estimate derived from either progenitor identifications in pre-explosion images, or from SN nebular spectroscopy, which arguably provide the most accurate estimates of \Minit. Our second selection criteria is that there must be MUSE data available of the host galaxy environment of our SN sample that is not contaminated by SN emission. Finally, in order to minimise the contribution from several stellar populations in our \ha\ EW measurement, we place a maximum limit of 150~pc on the effective spatial resolution of the MUSE data, which for a full-width half-maximum (FWHM) of 0.6 arcsec corresponds to a redshift $z<0.01$.

There are now around 45 ccSNe with progenitor \Minit\ estimates or limits from pre-explosion images, and a further $\sim 25$ with \Minit\ based on nebular spectral modelling. From this parent sample, $\sim 40$ are visible from the ESO Cerro Paranal Observatory, 14 of which already have publicly available archival MUSE data and have effective spatial resolution better than 150~pc. One of these is SN~1987A, which we do not include in our sample due to the continual bright emission from the SN remnant, and another, SN~2016gkg, is not included because the SN is still visible. Finally the host galaxy of ASASSN-14ha contains an AGN, emission from which is clearly visible at the SN position, contaminating our \ha\ EW measurement. This SN was therefore also excluded from our sample, leaving us with a final sample of 11 ccSNe, 10 of which are hydrogen-rich, type II SNe, and one which is a stripped-envelope, type Ib SN. There is a continuum in SN II properties between slow decliners (IIP) and fast decliners \citep[IIL; e.g.][]{agh+14}, but in this paper we will not distinguish between these subcategories. Details for our sample of SNe are given in Table~\ref{tab:sample}.

\begin{table*}
\caption{Details of the 11 ccSN included in our sample}\label{tab:sample}
\centering
\begin{threeparttable}
\begin{tabular}{llllccll}
\hline
\hline
SN & $z$ & RA & Dec & Type & E(B$-$V)$_{\rm Gal}^{\rm a}$ & Host galaxy & Distance \\
 & & & & & & (Mpc) \\
\hline
SN~1999br & 0.0033 & 13:00:41.8 & +02:29:45.8 & II & 0.02 & NGC 4900 & 14.7 \\
SN~2001X & 0.0049 & 15:21:55.5 & +05:03:42.1 & II & 0.03 & NGC 5921 & 21.7 \\
SN~2004dg & 0.0045 & 14:59:59.0 & +01:53:25.6 & II & 0.04 & NGC 5806 & 20.3 \\
SN~2006my & 0.0028 & 12:43:40.7 & +16:23:14.1 & II & 0.02 & NGC 4651 & 13.0 \\
SN~2008cn & 0.0086 & 12:40:55.6 & -40:58:12.1 & II & 0.14 & NGC 4603 & 38.1 \\
SN~2009H & 0.0044 & 02:45:58.4 & -07:35:00.3 & II & 0.02 & NGC 1084 & 19.0 \\
SN~2009N & 0.0034 & 12:31:09.5 & -08:02:56.3 & II & 0.02 & NGC 4487 & 14.3 \\
SN~2009md & 0.0043 & 10:48:26.3 & +12:32:02.8 & II & 0.02 & NGC 3389 & 19.2 \\
SN~2012P & 0.0045 & 14:59:59.1 & +01:53:24.3 & IIb & 0.04 & NGC 5806 & 20.1 \\
SN~2012ec & 0.0044 & 02:45:59.9 & -07:34:27.2 & II & 0.02 & NGC 1084 & 19.0 \\
iPTF13bvn & 0.0045 & 15:00:00.1 & +01:52:53.1 & Ib & 0.04 & NGC 5806 & 20.6 \\
\hline
\end{tabular}
\begin{tablenotes}
\small\item $^{\rm a}$ \citet{sf11}
\end{tablenotes}
\end{threeparttable}
\end{table*}

\section{Data reduction}
\label{sec:analysis}
Raw data were downloaded from the ESO archive together with the corresponding calibration files, and the data were reduced and combined into single MUSE data cubes for each individual SN in the sample using the standard ESO pipeline software version 1.2.1\footnotemark[1] \citep{wsu+14}. The pipeline applies corrections for bias level, flat-fields, illumination level, and geometric distortions. It then performs the wavelength calibration using daytime arclamp frames, which is then refined by skylines in the science data. We used the Zurich Atmospheric Purge \citep[ZAP; v2.0][]{slb+16} software package to subtract the sky lines from our data cubes, using an offset pointing to blank sky to measure the sky background when available. Finally we used the \texttt{molecfit} software package \citep{ssn+15} to correct for telluric absorption, using the spectrum of a star within the MUSE field of view to model the telluric absorption. The MUSE data cubes for all SNe in our sample had at least one star in the field of view. The star spectrum is fit with a physical model of the atmospheric molecular oxygen and water vapor content, using the three prominent telluric absorption bands within the MUSE wavelength range (centered at around 6870~\AA, 7600~\AA, and 7630\AA) to constrain the fit. The best-fit telluric absorption for the single star is then applied to all spaxels within the MUSE data cube. 
\footnotetext[1]{http://www.eso.org/sci/software/pipelines/}

The next step in our analysis was to separate the stellar and gas-phase components of the galaxies in our sample, in order to accurately measure the gas emission line fluxes. Hydrogen Balmer emission lines are strongly affected by absorption from the atmospheres of stars, in particular \hb\ and higher order lines along the Balmer series, and they become increasingly pronounced for older stars. If not taken into account, this stellar absorption can inadvertently affect science results, most notably overestimating the Balmer decrement, and thus overestimating the amount of dust extinction. This in turn will affect physical quantities that rely on dust corrections, such as gas phase oxygen abundance and star formation rate. To model the stellar continuum, we followed the same procedure as described in \citet{kks+17}, in which we extract spectra from regions $3\times 3$ spaxels in size, and use {\sc starlight} \citep{cms+05,csg+09} to fit a linear superposition of template spectra based on the \citet{bc03} stellar population models. We then linearly scale our best-fit stellar template by the fraction of light in each of the spaxels in the $3\times 3$ region, and subtract the stellar component from the original data cube to produce a gas phase-only emission data cube. Our highest spatial resolution data was 0.6 arcsec in pixel units, which is comparable to the scale of the regions that we used to separate the stellar and gas emission components. 

\begin{figure}
\centering
  \includegraphics[width=1.0\linewidth]{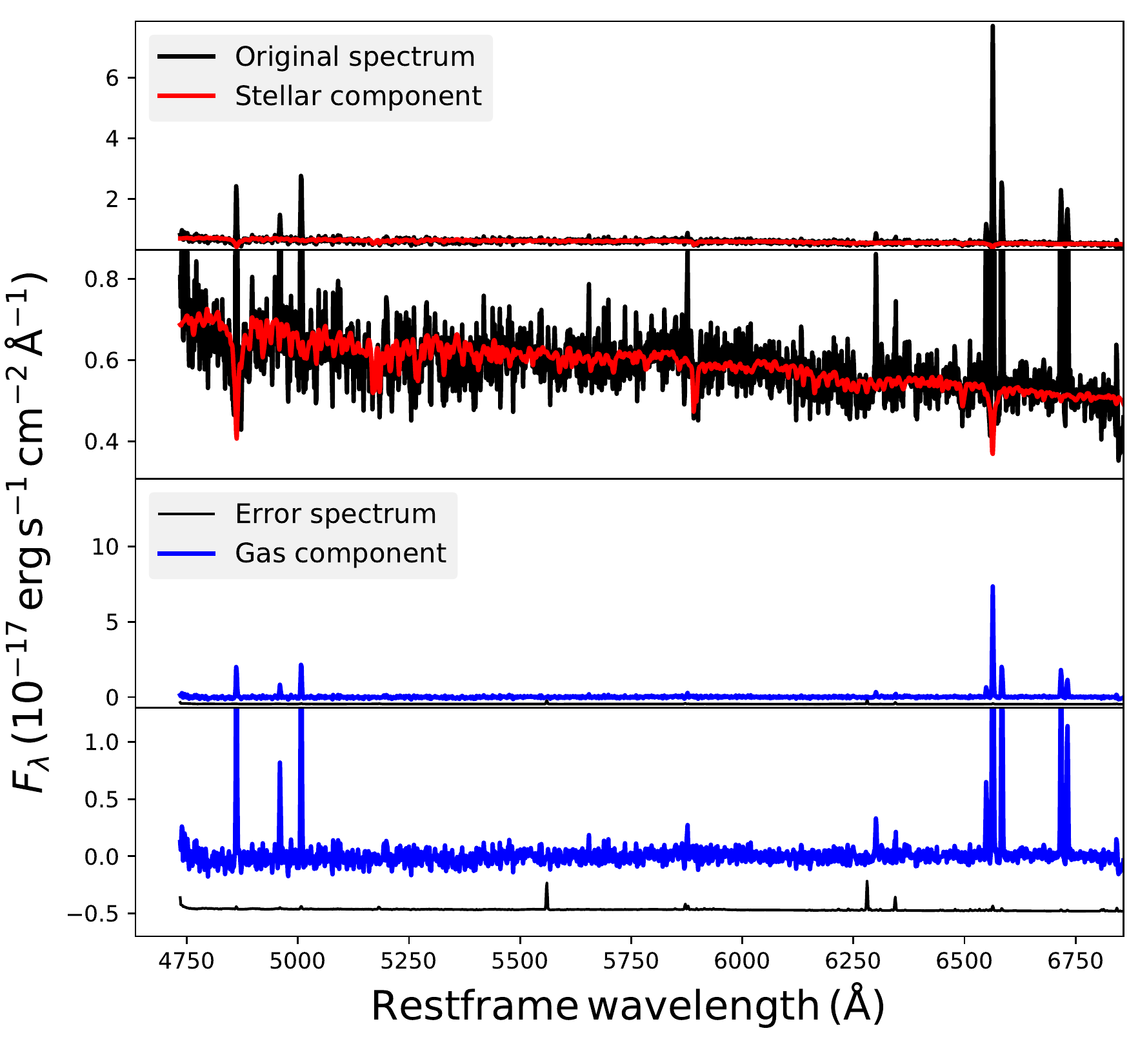}
  \caption{Example spectra taken at the location of SN~1999br to illustrate the quality of our {\sc starlight} fits and our procedure to separate stellar and gas-phase components. {\em Top}: original spectrum (black) extracted at the SN position with the best-fit {\sc starlight} stellar model overplotted (red). {\em Second}: zoom-in of the continuum. {\em Third}: the resulting spectrum of the gas-phase component (blue) after removing the model from the observed spectrum shown in the top panel. {\em Bottom}: zoom-in of the gas-phase contribution (blue) together with the error spectrum (black).}\label{fig:slfits}
\end{figure}

To illustrate the quality of our starlight fits, in Fig.~\ref{fig:slfits} we show the total emission spectrum extracted at the position of SN~1999br with the best fit {\sc starlight} model overplotted (top two panels), as well as the subsequent gas phase-only emission spectrum after subtracting the model stellar component (bottom two panels). The Balmer lines can be clearly seen in the total emission spectrum (black line) in absorption and emission (especially \hb\ at $\sim 4880$\AA), and the {\sc starlight} model (red curve) re-produces well these stellar absorption features and the overall continuum. The effectiveness of our method to remove the stellar emission component can be seen in the lack of absorption around the Balmer lines in the gas-phase only emission spectrum plotted (bottom two panels). Nevertheless, despite the good quality of the stellar continuum fits, the physical parameters associated to the best-fit {\sc starlight} models are unlikely to be reliable due to degeneracies that exist between parameters, such as stellar metallicity and age. However, this does not affect our analysis, since we are only interested in reproducing the stellar continuum, and not in the physical properties themselves that produce the model continuum.

\subsection{Astrometry and SN position}
\label{ssec:SNpos}
Stars within young stellar clusters with ages $\la 10$~Myr remain associated out to scales of $\sim 100$~pc \citep[e.g.][]{gp97,lkb+11,cro13,whm+18}, which for the redshifts of our SN sample ($z$=0.0017$-$0.0086), corresponds to 0.8$-$4.3 arcsec 3 in our MUSE data cubes. In order to minimise the contamination within our SN aperture from emission from other, unrelated stellar populations, it is therefore important to locate the SN explosion site within our MUSE data cubes to high precision. For 4 of the 11 sources in our sample, the SN was observed with {\em HST}, enabling us to locate the position of the SN within the {\em HST} images from the standard pipeline-reduced archival data to within 0.2 arcsec. To locate the relative SN position in our MUSE data cubes, we used the GAIA catalogue to shift the {\em HST} and MUSE data to the same astrometric reference frame. This resulted in typical shifts in our MUSE astrometry of 0.1$-$0.2 arcsec in both directions. Due to the comparatively large MUSE field of view, we were always able to locate at least two foreground stars within our MUSE data cubes, allowing us to tie down our MUSE astrometry to the {\em Gaia} catalogue to better than 50~mas, which is within a MUSE spaxel (0.2 arcsec).

A further 3 SNe in our sample with no {\em HST} observations have reported accurate offsets between the SN position and nearby stars, allowing us to again locate the position of the SN within our MUSE datacubes to within one spaxel. For the remaining 4 SNe with no {\em HST} data or accurate relative offsets, we used the SN positions relative to the host galaxy nucleus as provided in the Open Supernova Catalogue\footnotemark[2] to locate the SN within our {\em Gaia}-aligned MUSE data cubes \citep{gpk+17}.
\footnotetext[2]{https://sne.space/}

With accurate SN positions in place, we then measured some key gas-phase properties at the SN location within the effective spatial resolution of our data. This is given by the FWHM of the point spread function, measured from stars in the MUSE field of view. This ranged between 0.6 arcsec and 1.0 arcsec, corresponding to physical scales of 50$-$125~pc. The FWHM, the physical spatial resolution of our data, and a number of physical properties associated to the ionised gas at the location of the SNe in our sample, described in the next section, are listed in Table~\ref{tab:prop}.

\begin{table*}
\centering
\caption{Physical properties inferred from MUSE IFU spectroscopy at the position of the ccSNe in our sample. \label{tab:prop}}
\begin{threeparttable}
\begin{tabular}{lcccccc}
\hline
\hline\noalign{\smallskip}
{SN} & FWHM & Phys. scale & {\ha\ EW} & {E(B$-$V)$_{\rm host}$} & {12+log(O/H)} & n$_e$\\
 & (arcsec) & (pc) & (\AA) & & (PP04 O3N2) & (cm$^{-3}$) \\
\hline\noalign{\smallskip}
SN~1999br & 0.8 & 55 & $50.5\pm 0.7$ & $0.14$ & $8.54\pm 0.11$ & $<100$ \\
SN~2001X	 & 0.8 & 82 & $22.8\pm 0.1$ & $0.18$ & $8.83\pm 0.30$ & $<100$ \\
SN~2004dg & 0.6 & 56 & $106.6\pm 0.8$ & $0.32$ & $8.87\pm 0.21$ & $<100$ \\
SN~2006my & 0.9 & 51 & $13.5\pm 0.1$ & $0.00$ & $8.71\pm 0.21$ & $<100$ \\
SN~2008cn & 0.7 & 125 & $6.5\pm 0.1$ & $0.09$ & $-$ & $<100$ \\
SN~2008cn$^{\rm a}$ & $-$ & $-$ & $12.2\pm 0.1$ & 0.13 & $8.66\pm 0.36$ & $<100$ \\
SN~2009H & 0.9 & 82 & $56.4\pm 0.1$ & $0.58$ & $8.71\pm 0.10$ & $<100$ \\
SN~2009N & 1.0 & 71 & $6.5\pm 0.1$ & $-$ & $-$ & $<100$ \\
SN~2009N$^{\rm a}$ & $-$ & $-$ & $41.8\pm 0.4$ & 0.23 & $8.66\pm 0.15$ & $187\pm 55$ \\
SN~2009md & 0.6 & 53 & $11.8\pm 0.1$ & 0.12 & $8.60\pm 0.23$ & $184\pm 29$\\
SN~2012P & 0.6 & 56 & $114.7\pm 0.2$ & 0.42 & $8.88\pm 0.05$ & $<100$ \\
SN~2012ec & 0.9 & 82 & $26.4\pm 0.1$ & 0.50 & $8.67\pm 0.09$ & $<100$ \\
iPTF13bvn & 0.6 & 56 & $13.6\pm 0.1$ & 0.20 & $8.83\pm 0.36$ & $134\pm 10$ \\
\hline
\end{tabular}
\begin{tablenotes}
\small\item $^{\rm a}$ Analysis done at region nearby to the SN position which has \ha\ EW $>10$ \AA. 
\end{tablenotes}
\end{threeparttable}
\end{table*}

\section{Analysis and results}\label{sec:results}
\subsection{\ha\ equivalent width}
\label{ssec:haew}
To measure the \ha\ EW at the SN position, we extract a stacked spectrum from the gas plus stellar emission data cube, taken within a circular region centred at the SN position and with a diameter equal to the image FWHM. We then fit the stellar continuum and \ha\ emission line using a Gaussian plus constant to measure the \ha\ EW. Zoom-ins of the maps around each of the SNe in our sample are shown in Fig~\ref{fig:env} (left most panels). In this figure we also show zoom-ins of the stellar continuum centred on the $R$-band (middle panel) to illustrate the contribution to the \ha\ EW from stellar light. Note that SN~2004dg and SN~2012P occurred within the same region of the same host galaxy, and they are therefore together in Fig.~\ref{fig:env} (third row). There is quite a spread in \ha\ EW measured at the location of our SN sample. For example, SN~2008cn and SN~2009N are far from any young \hii\ regions, whereas SN~2004dg and SN~2012P are located on a bright \hii\ region (see Fig~\ref{fig:env}, left and middle panels). However, the SNe generally lie in regions of their host galaxy with \ha\ EW $\la 50$~\AA\ (see Table~\ref{tab:prop}), consistent with the general disassociation between SN II and star formation \citep[e.g.][]{ahj+12,gsm+14,kag+18}.

An important assumption made when using the \ha\ EW and other emission lines to infer properties on the stellar population at the SN position is that this line emission arises from excitation by stellar light. At low \ha\ EWs this assumption may no longer be valid, and the emission may instead be dominated by diffuse gas emission (DIG). The origin of DIG emission is still not clear, but some insight is given by the location of DIG emission on the Baldwin, Phillips and Terlevich (BPT) diagram \citep{bpt81}, which provides a diagnostic to identify different excitation mechanisms that may ionise gas. The original BPT diagram shows the ratio of [\ion{O}{iii}]$\lambda 5007$/\hb\ against [\ion{N}{ii}]$\lambda$6584/\ha, in which gas ionised by star formation, AGN activity, or low ionisation emission line regions (LIERs) occupies distinct regions of the diagram. Spectroscopic studies show that DIG emission lies in the LIERs portion of the BPT diagram \citep[e.g.][]{mrh06,rd03,kjk+16,zyb+17}.

\begin{figure}
\includegraphics[width=1.1\linewidth]{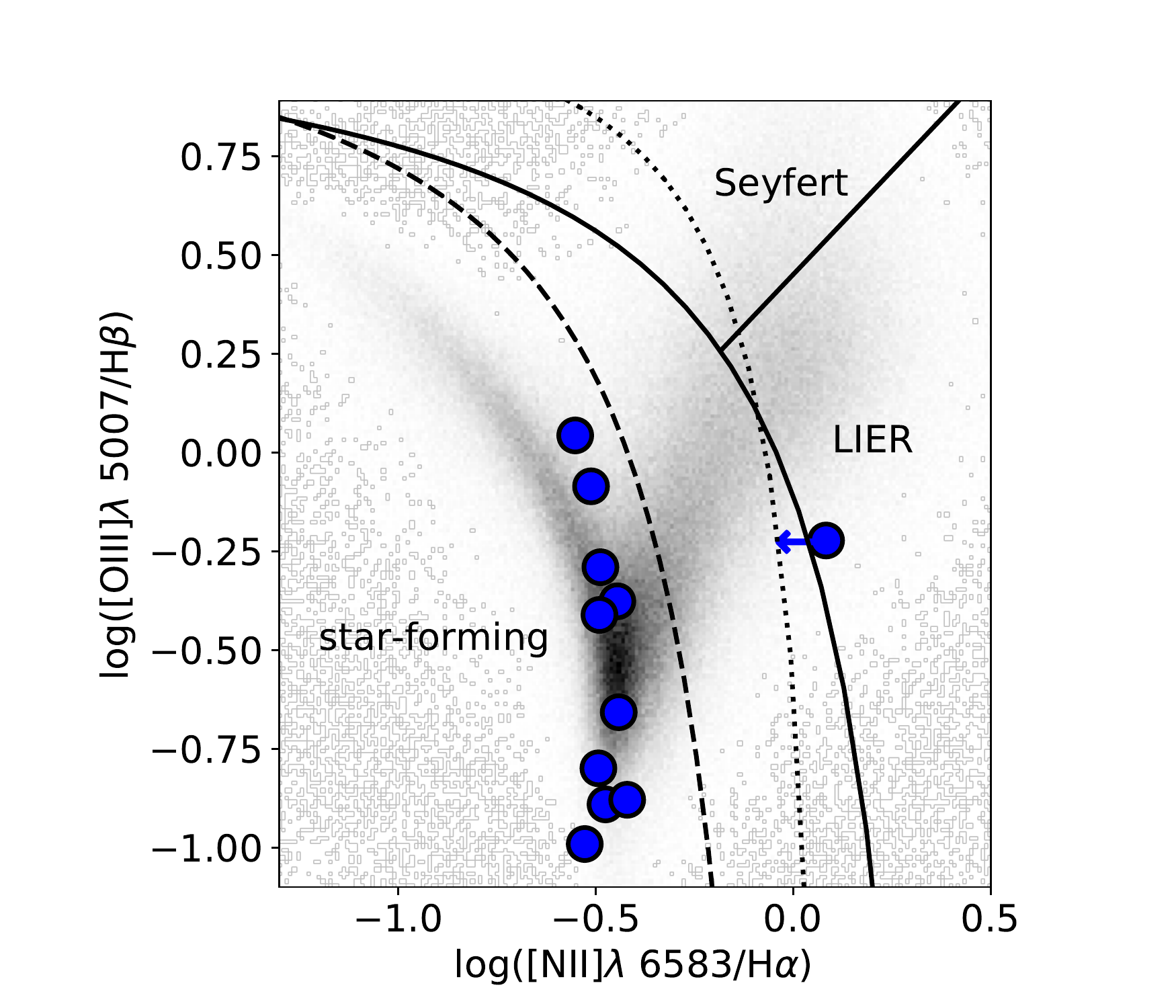}
\caption{BPT diagram with population of SDSS galaxies shown in grey scale, and the emission at the location of our SN sample represented by the blue circles. The data point with a leftward pointing arrow corresponds to SN~2009N. The dashed, solid and dotted lines are the demarcation lines that separate gas ionised by stellar light and AGN activity according to the prescriptions from \citet{kht+03}, \citet{kds+01} and \citet{xse18}, respectively. The solid line in the top right corner of the plot separates Seyfert and LIER emission according to \citep{sts+07}.}\label{fig:BPT}
\end{figure}

In order to check whether the ionised gas at the SN position likely originates from excitation from stars or from some other mechanism, we plot the line ratios at the SN location on the BPT diagram in Fig~\ref{fig:BPT}. All emission lines are measured from Gaussian fits to the gas phase spectrum extracted at the SN position. The distribution of SDSS galaxies are plotted in grey scale, and the red circles show the emission line ratios at the positions of our SNe. In general the ionised gas properties in the vicinity of our SNe trace the distribution of SDSS star forming galaxies on the BPT diagram, with the exception of SN~2009N, where only an upper limit of [\ion{N}{ii}]$\lambda 6584<0.22\times 10^{-17}$~erg~s~cm$^{-2}$ can be placed at the location of the SN. Not including this unconstrained measurement, the gas properties at the position of the SNe are all within the `star-forming' region of the diagram according to the demarcation lines between stellar and AGN excitation from  \citet{kht+03}, \citet{kds+01} and \citet{xse18}, represented by the dashed, solid and dotted lines respectively. The solid line in the top right of the plot shows the separation between Seyfert and LIER emission according to \cite{sts+07}. Independent analysis on the properties of DIG emission also places a limit of \ha\ EW$<3$\AA, above which DIG emission should not dominate \citep{csm+11,bmt+17}. This is smaller than the \ha\ EWs that we measure at the position of our SN sample, where the lowest values correspond to SN~2008cn and SN~2009N, both of which lie in regions of their host galaxy with \ha\ EW$\sim 6.5$\AA. These factors combined suggest that the ionised gas at the location of our SNe (apart from perhaps SN~2009N) is ionised by star formation, and we can therefore use these data to study the stellar population at the SN position.

\begin{figure*}
\centering
  \includegraphics[width=1.0\linewidth]{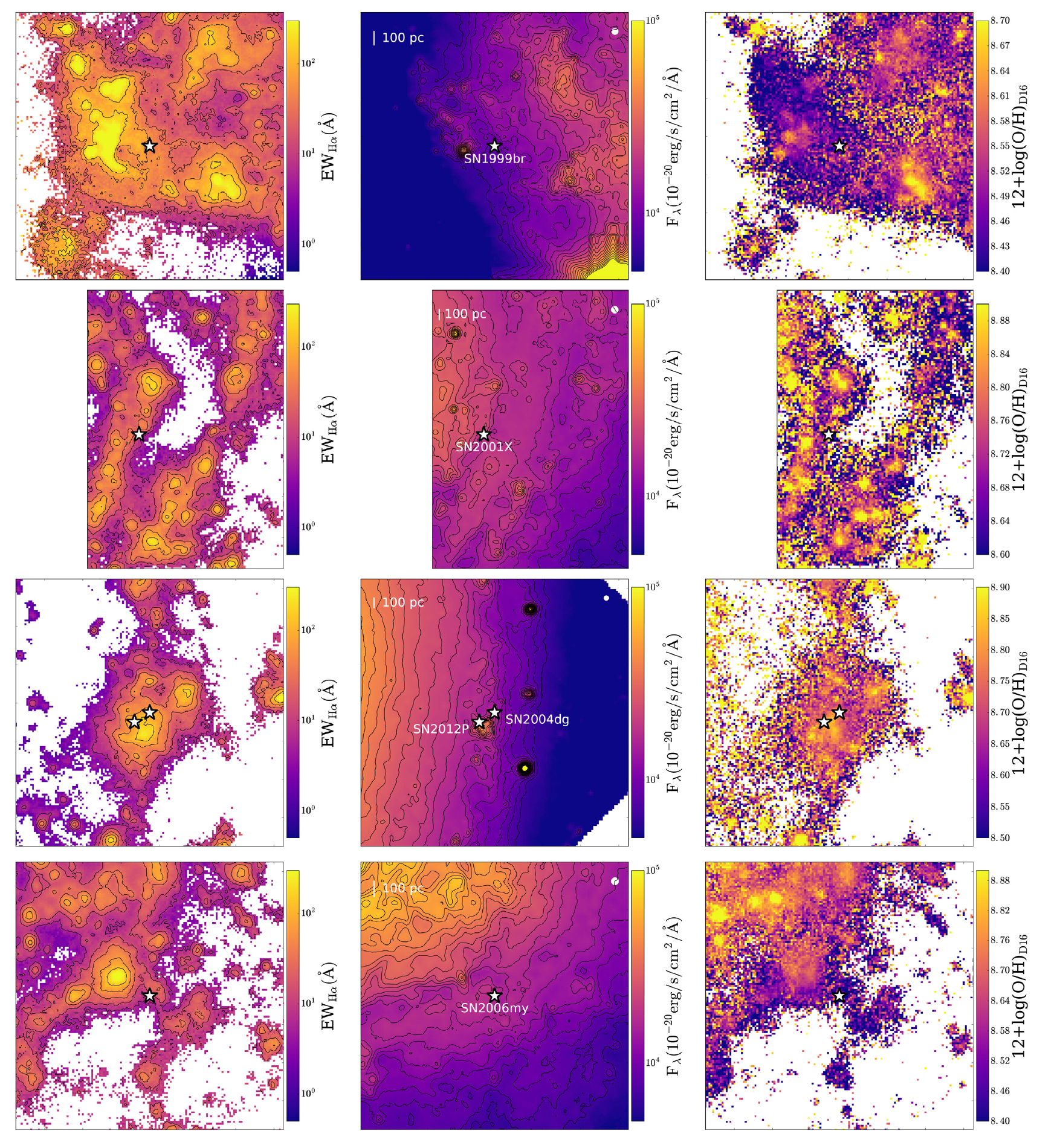}
\caption{Zoom-ins around the locations of the 11 SNe in our sample of maps showing the \ha\ EW (left), stellar continuum emission centred on the $R$-band (middle), and gas-phase metallicity based on the D16 diagnostic \citep{dks+16} (right). Each row corresponds to a different SN apart from the third row, which shows SN~2004dg and SN~2012P together, since they both occurred within the same region of the same host galaxy. In the middle panel we give the SN name, figure scale (top left) and the PSF of the data (top right). In each plot the position of the SN is indicated by a white solid star. In the two cases where the \ha\ EW measured at the SN position was $<10$\AA\ (SN~2008cn and SN~2009N), a white cross in the \ha\ EW maps indicates the location of the nearest region to the SN with \ha\ EW$>10$\AA\ within an area that is a FWHM of the data. Pixels with a signal-to-noise ratio $<3$ are left white.}\label{fig:env}
\end{figure*}

\begin{figure*}
\centering
\setcounter{figure}{1}
  \includegraphics[width=1.0\linewidth]{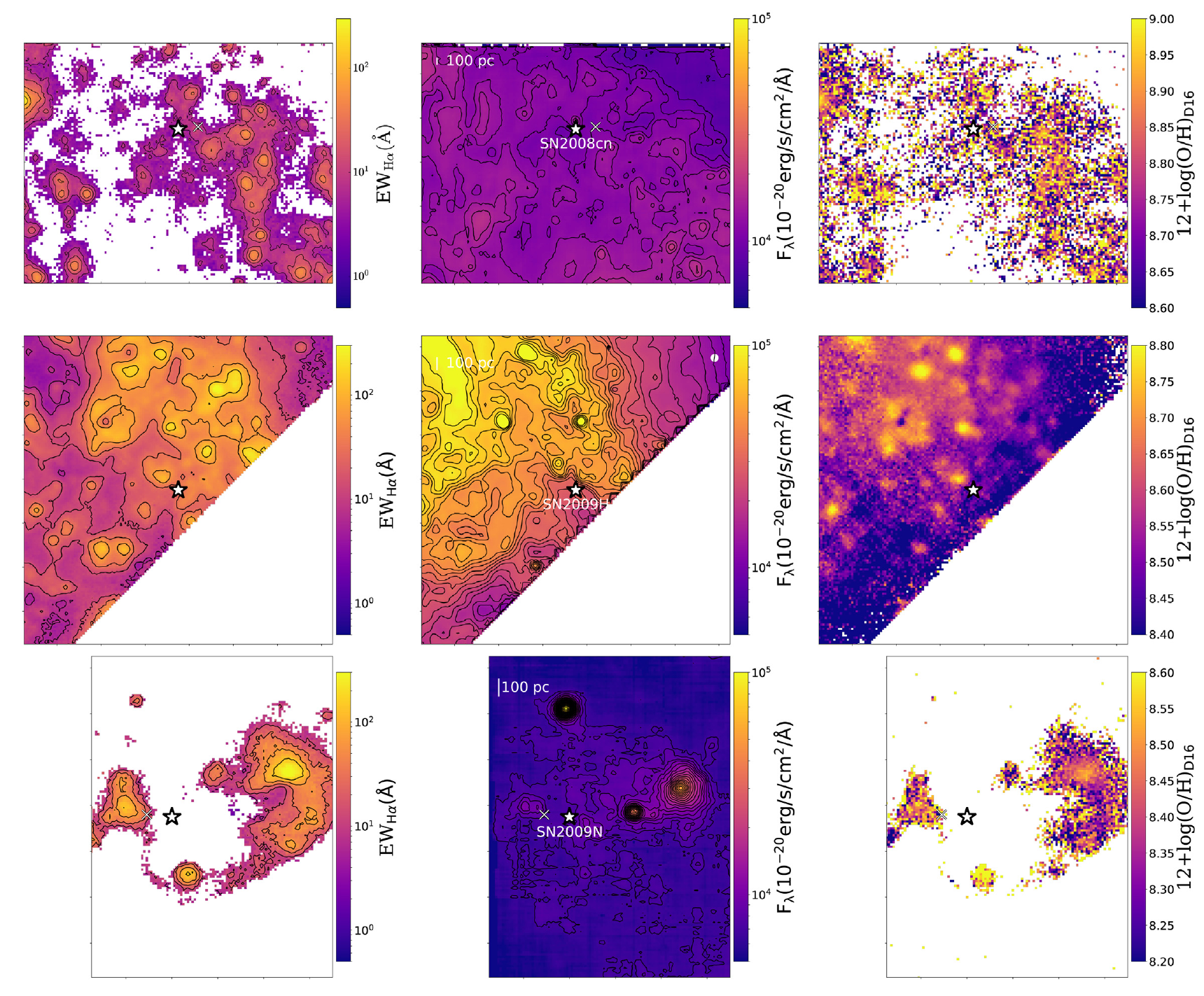}
\caption{cont.}
\end{figure*}

\begin{figure*}
\centering
\setcounter{figure}{1}
  \includegraphics[width=1.0\linewidth]{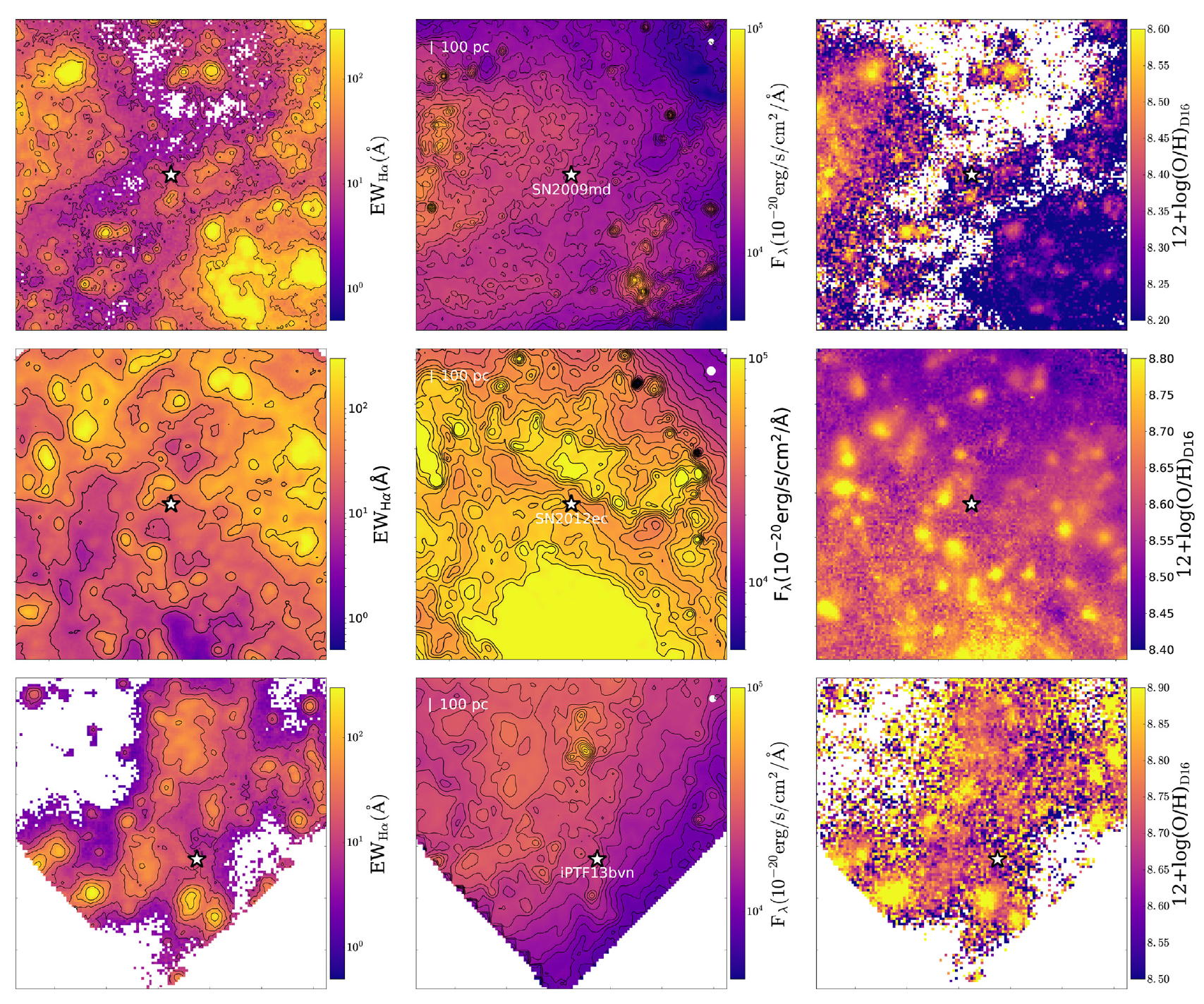}
\caption{cont.}
\end{figure*}

\subsection{Stellar population synthesis models}
\label{ssec:bpass}
There are a number of stellar population synthesis (SPS) models that have been developed to interpret and extract physical properties from galaxy observations, with the most sophisticated including emission from stars and ionised gas, and may also take into account the effects of intervening dust. When considering the most standard SPS models, they show relatively good agreement in the relation between the equivalent width of hydrogen recombination lines and stellar age \citep[e.g.][]{kga+16}, but models diverge as they become more complex. For example, a problem with the most standard SPS models is that they fail to reproduce the bluer spectral properties observed in star forming galaxies, producing insufficient far-UV flux, even when assuming a continuous star formation history \citep[e.g.][]{lkl10}. This in itself is not surprising given the overwhelming observational evidence that a significant fraction of stars undergo binary interactions during their lifetime \citep[e.g.][]{dk13,sll+14,md17}. Indeed, improvements have been found in population models when including the effects of stellar rotation \citep[e.g.][]{lle+12} or binary system interactions \citep[][]{dv03,es09,di17,zdj+19}, which have the effect of prolonging the presence of ionizing photons from a stellar population beyond that expected for a single star population. These models correspondingly also produce larger Balmer line EWs for the same given stellar population age.

In this work we have chosen to use the Binary Population and Spectral Synthesis models \citep[BPASS;][]{es09,es12} to infer the stellar population age from the nearby environment emission line spectra due to their relative success at matching observations, and their ease of implementation. It is thought that up to 70\% of massive stars will undergo mass transfer with a binary companion during their lifetime \citep{sdd+12}, which will alter the star's structure and evolution quite significantly. We use BPASS v2.1 SPS models\footnotemark[3] \citep{esx+17}, which were publicly released in October 2017 \citep{esx+17} and also include synthetic spectra of gas ionised by the BPASS stellar populations. These models are created by sampling from an initial mass function (IMF) and from a range of binary system properties to create composite stellar spectra at a range of ages, from 1~Myr to 10~Gyr in logarithmic steps of 0.1~dex. Although a more recent version of BPASS models was released in May 2018 \citep[v2.2.1;][]{se18}, we chose to use v2.1 because this data release includes synthetic spectra of gas excited by the model stellar populations. These synthetic spectra were created using the radiative transfer code {\sc cloudy} v13.03 \citep{fkv+98,fpv+13} with BPASS v2.1 models as input ionisation spectra \citep{xse18}. The interstellar gas was assumed to have the same metallicity and abundance patterns as the stars, and models are available for 13 different metallicities spanning the range $Z=10^{-5}$ to $Z=0.040$, where $Z=0.014$ corresponds to roughly solar. Models have also been created for a range of gas hydrogen densities, between 1$-$1000~cm$^{-3}$ in logarithmic intervals of 0.5~dex, and for 21 values of ionisation parameter between $\log U=-3.5$ and $\log U=-1.5$. This produces over 40,000 SPS models for the single and the binary interaction SPS models.
\footnotetext[3]{Available to download at http://bpass.auckland.ac.nz}

The greatest difference between v2.1 and v2.2 for the purposes of this paper is an increase in the fraction of single stars in the latter case. When only considering the observed \ha\ EW, BPASS models v2.2 would therefore yield systematically younger stellar population ages than the v2.1 models. However, in our analysis, we fit the binary system BPASS models to ionised gas emission line ratios as well as \ha\ EW (section~\ref{ssec:fits}), which circumvents this effect. Line ratios are predominantly set by the temperature of the hottest stars in the population, which in the binary case are the helium cores revealed by binary interactions. The observed line ratios are thus relatively independent of binary fraction. Furthermore, we also investigate the effect of `photon leakage' (section~\ref{sssec:leakybinfits}), which has a similar effect to changing the stellar binary fraction. Therefore, although we may expect BPASS v2.2 models to yield somewhat younger stellar ages, we don't expect differences to be significant to the results presented in sections~\ref{sssec:leakybinfits} and \ref{sssec:binfits}.

The synthetic spectra are available as ascii files, and contain the model flux and equivalent widths of 21 emission lines spanning the UV and optical wavelength range in tabular form, making it relatively easy to directly compare the model emission line properties to observations. Before reaching this step, we first select a subset of the BPASS SPS models to compare to our SN data based on the input model gas metallicity and hydrogen density, which are properties that we can measure using line ratio diagnostics.

\begin{figure*}
  \includegraphics[width=1.0\linewidth]{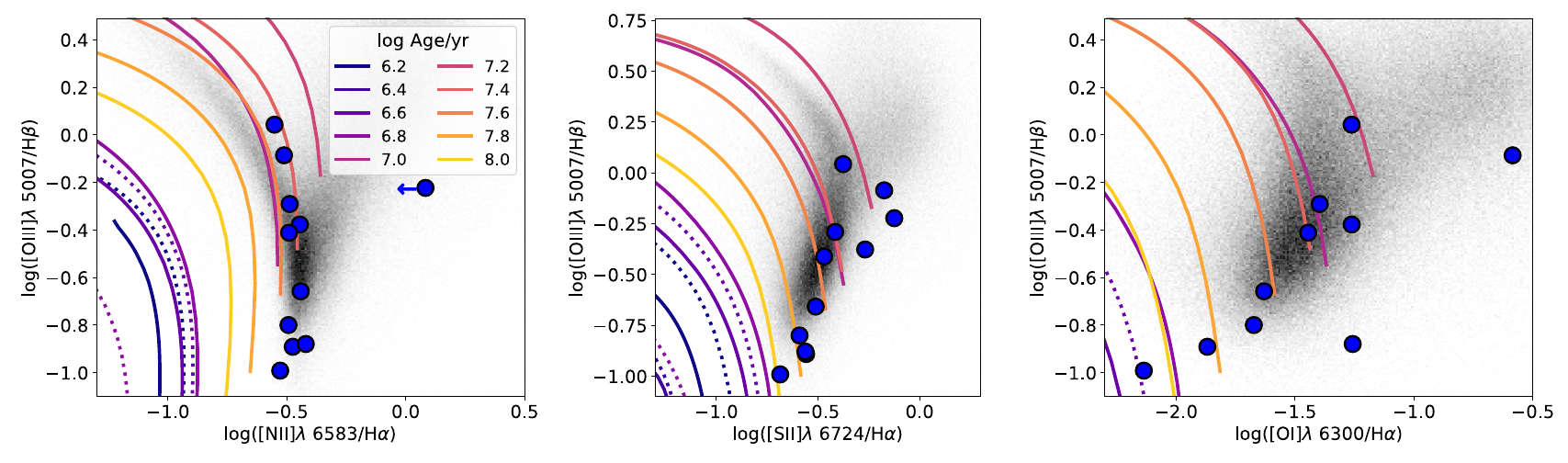}
\caption{Three BPT diagnostic diagrams with blue circles representing the interstellar gas properties at the location of the SN in our sample. The $x$-axis in each panel differs, showing the relation between $\log([\ion{O}{iii}]/\lambda 5007)$/\hb\ against $\log([\ion{N}{ii}]/\lambda 6583)$/H$\alpha$, $\log([\ion{S}{ii}]/\lambda 6724)$/H$\alpha$ and $\log([\ion{O}{i}]/\lambda 6300)$/H$\alpha$ in the left, middle and right panels respectively. Single star (dotted lines) and binary system (solid lines) BPASS models are also overlaid for models with metallicity $Z=0.014$ (corresponding to oxygen abundance $12+\log(\mathrm{O/H})\sim 8.7$) and hydrogen density $\log\mathrm{(n_H)}=1.5$, for a number of stellar population ages given in the legend of the first panel. The BPASS tracks trace the position of the models on the BPT diagram for decreasing ionisation parameters, $U$, starting from $\log(U)=-1.5$ (upper left) down to $\log(U)=-3.5$ (lower right).}\label{fig:allBPTs}
\end{figure*}

\subsection{Gas-phase oxygen abundance and hydrogen density}
\label{ssec:oh}
An important physical property that affects how the measured \ha\ EW relates to the age of the ionising stars is the stellar and gas-phase metallicity. Although with our MUSE data it is difficult to determine the metallicity of the stars themselves, we can estimate the gas-phase oxygen abundance, which since oxygen is the most abundant metal, is commonly used as a proxy for metallicity. We then assume that the gas phase metallicity is the same as the metallicity of the stars, as is also done in the BPASS SPS models.

Many diagnostics exist to determine the gas-phase oxygen abundance of ionised gas, with the so-called strong-line diagnostics being the most widely used due to their applicability in more metal-rich and high-z galaxies. Other, albeit more accurate diagnostics, rely on weaker emission lines, which are challenging to detect. Large systematic differences in the oxygen abundance are known to exist between strong-line metallicity diagnostics \cite[e.g][]{ke08}, and there is a wealth of literature on the origins of these differences and on the strengths and weaknesses of the various metallicity diagnostics available \citep[e.g.][]{kd02,esp+06,ykg12,am13,kks+17}. For our sample, which is exclusively at $z<0.01$, the MUSE data do not probe the [\ion{O}{ii}]$\lambda\lambda 3726,3729$ doublet, limiting our choice of metallicity diagnostics. Of the metallicity diagnostics that can be applied to our data, the \citet{pp04} O3N2 and N2 are the most widely used, as well as the updated O3N2 and N2 calibrations presented in \citet{mrs+13}. These diagnostics work on the basis that there exists an anti-correlation between ionisation parameter (fraction of ionizing photons per hydrogen number density) and oxygen abundance. Although this anti-correlation can be seen in samples of \hii\ regions, there is evidence that the relation breaks down at small scales \citep[e.g.][]{kks+17}. Resolved spectral observations show that \hii\ regions have highest ionisation parameters at the centres, which decrease towards the edges, such that O3N2 and N2 diagnostics produce inverted metallicity profiles that are lowest at the centre, contrary to expectation \citep{rmv+10,mrk+11,kks+17,mlk+18}.

A more recently proposed metallicity diagnostic by \citet{dks+16} uses the [\ion{S}{ii}] line to constrain the gas ionisation parameter, which produces galaxy metallicity distributions more in line with expectations \citep[i.e. the centres of \hii\ regions are more chemically enriched than their edges;][]{kks+17}. Moreover, when applying a large number of metallicity diagnostics to the BPASS synthetic spectra, \citet{xse18} found the \citet{dks+16} diagnostic (N2S2 from now on) to match their input model metallicities more closely than the \citet{pp04} and \citet{mrs+13} O3N2 and N2 diagnostics. In Fig.~\ref{fig:env}, right-most column, we therefore show the metallicity distribution in the vicinity of the SN in our sample based on the N2S2 metallicity diagnostic.

Similar to our \ha\ EW measurements, the metallicity at the SN location for any given diagnostic was measured from a stacked spectrum extracted within a circular region centred at the SN position from the gas-only data cube. When applying the N2S2 metallicity diagnostic, we found that for 5 of the 11 SNe in our sample, the measured metallicity was $12+\log\rm{(O/H)}>9.05$ \citep{dks+16}, where the N2S2 diagnostic is no longer valid. Therefore, despite some of the benefits in the N2S2 discussed above, we chose to use the widely applied O3N2 diagnostic in our subsequent analysis in order to have a metallicity value for the entire sample that is consistently derived using the same diagnostic. The \citet{pp04} O3N2 metallicities that we measure at the SN positions are listed in Table~\ref{tab:prop}, but given the variation between different metallicity diagnostics, in Table~\ref{tab:Zall}, we also list the metallicities that we measure when using the N2S2, and N2 metallicity diagnostics, as well as the results from the re-calibrated O3N2 and N2 diagnostics from \citet{mrs+13}. All metallicities have been corrected for host galaxy and Milky Way dust extinction. To correct for Galactic extinction we used the $E(B-V)$ reddening values from \citet{sf11} and assumed a CCM extinction law \citep{ccm89} using the average total-to-selective Galactic extinction value $R_V=3.08$. The host galaxy dust reddening, E(B$-$V)$_{\rm host}$, was determined from the \ha/\hb\ Balmer decrement, which was compared to an intrinsic ratio of 2.86 appropriate for an ISM temperature of $T=10^4$~K and an electron density of $n_e=10^2-10^4$~cm$^{-3}$ \citep{ost89}. The corresponding dust extinction at the relevant emission line wavelengths was derived assuming a Calzetti dust attenuation law \citep{cab+00}. All emission line fluxes used to derive gas-phase metallicities are given in Table~\ref{tab:emlines}, uncorrected for dust extinction, where we also provide the host galaxy dust reddening at the SN positions based on the Balmer decrement. 

In the case of SN~2008cn and SN~2009N, the emission lines at the SN position are weak, as already highlighted in section~\ref{ssec:haew}, and the corresponding gas-phase metallicity is thus unconstrained. In order to get some idea of the gas-phase metallicity in the vicinity of these SNe, we extract a spectrum from the gas-phase cube in a region nearby to the SN with \ha\ EW$>10$\AA, where the emission lines are sufficiently strong to be able to measure a metallicity. These nearby regions are indicated in the relevant panels of Fig.~\ref{fig:env} with a white cross.

We also use the extracted, gas-phase spectra to determine the gas hydrogen density at the location of the SN, where we use the relation between electron density and the sulphur [\ion{S}{ii}]$\lambda 6717/\lambda 6731$ line ratio provided in \citet{of06}, which is appropriate for densities in the range n$_e=10^2-10^4$~cm$^{-3}$. In most cases we find that the measured gas density is $<10^2$~cm$^{-3}$, allowing us to only place an upper limit on this parameter. In Table~\ref{tab:prop} we give the O3N2 metallicity, the galaxy E(B-V) determined using the Balmer decrement, and $n_e$ (or limit) at the location of the SNe as well as in nearby regions where relevant.

\subsection{Stellar age from diagnostic emission line ratios}
\label{ssec:fits}
To find the most appropriate model that best describes our data, we first select a subset of the single star and the binary system evolution BPASS SPS models with metallicities and gas densities comparable to the properties measured at the SN positions (section \ref{ssec:oh}). These observational constraints typically reduce the number of SPS models to $< 1000$, depending on how accurately we are able to measure the gas phase metallicity. To select the model from this subset of templates that most closely represents our data, we follow a similar approach for the single star and the binary system SPS models, comparing our measured emission line properties to model templates, but using a larger number of observables in the latter case. We describe the procedure that we follow for the single and binary model analysis below.

\subsubsection{Binary evolution models}
\label{sssec:binfits}
In the latter case, we follow a similar approach to \citet{xse18}, who made use of the additional information that is available on the ionising stellar population from the diagnostic line ratios [\ion{O}{iii}]$\lambda 5007$/\hb, [\ion{N}{ii}]$\lambda$6584/\ha, [\ion{S}{ii}]$\lambda\lambda$6717,6731//\ha, and [\ion{O}{i}$]\lambda$6300/\ha. An example of how these line ratios change in the BPASS models with stellar age, and between the single and binary evolution models is illustrated in Fig.~\ref{fig:allBPTs}. In this figure we plot a subset of single and binary system BPASS models on the three diagnostic BPT diagrams that show the relation between [\ion{O}{iii}]$\lambda 5007$/\hb\ against [\ion{N}{ii}]$\lambda$6584/\ha, [\ion{S}{ii}]$\lambda\lambda$6717,6731//\ha, and [\ion{O}{i}$]\lambda$6300/\ha. Similar to Fig.~\ref{fig:BPT}, the blue data points represent the ionised gas properties at the SN locations, and the distribution of SDSS galaxies is plotted in grey scale. The selection of models shown correspond to the BPASS single (solid) and binary (dashed) models with metallicities Z=0.014, corresponding to an oxygen abundance of $12+\log(\mathrm{O/H})\sim 8.7$, and a hydrogen gas density of $\log(\mathrm{n_H}/\mathrm{cm}^{-3})=1.5$, which are the mean metallicity and gas density that we measure at the position of our SN sample. All emission line fluxes used in the three, diagnostic BPT diagrams are given in Table~\ref{tab:emlines}, uncorrected for dust extinction.

In our analysis, we compare the \ha\ EW and the relevant line ratios measured at the SN position to the BPASS model template predictions, where we only consider those models that have metallicities within 0.2 dex of our O3N2 best-fit value (corresponding to the systematic uncertainty on O3N2), and an electron column density, $n_e$ consistent with our measured values or limits (see Table~\ref{tab:prop}). In order to identify the best-matched model template, for each model in our subset we calculate the sum of the difference between the measured and model \ha\ EW and relevant line ratios, weighted by the measured uncertainties, giving us a chi-squared. We then select the model which minimises the chi-squared, and to see how degenerate the model templates are, we also identify all models with a chi-squared value that is within 10\% of the best-matched model. We found that in most cases, models within 10\% of the best-matched model are all of the same stellar age. The exception to this was SN~2012P, where models with stellar ages in the range 12.6$-$15.8~Myr were similarly well matched to the observations.

A concern in our analysis was that we found that the best-matched models generally had line ratios that differed significantly from the observations. We illustrate this in Fig.~\ref{fig:SNBPTs}, where we plot a subset of our SN on the [\ion{S}{ii}]$\lambda$6717,6731/\ha\ BPT diagram with three sets of BPASS binary evolution curves overlaid, corresponding to models with $\log(\mathrm{n_H}/\mathrm{cm}^{-3})=1.5$ and metallicities of Z=0.008, 0.010 and 0.020 respectively. In each panel, we show a subset of the SNe in our sample with environmental conditions best-matched to the set of plotted BPASS models according to the analysis described above, where the SN data points are colour-coded according to the best-matched age. This figure shows that the SN data points are often not the same colour as the model curve closest to them (i.e. same age), the latter often being of a correspondingly younger age, and the data points can in fact often lie far from the model curve of the same corresponding colour. This suggests that the best-matched stellar ages are being largely driven by the \ha\ EW value.

\begin{table*}
\centering
\caption{Estimated stellar population age based on our analysis on the circumstellar emission lines at the SN position for both single star and binary system evolution models, in the latter case with and without considering photon leakage.\label{tab:SNage}}
\begin{threeparttable}
\begin{tabular}{l|cc|cc|cc|c|c}
\hline
\hline\noalign{\smallskip}
{SN} & \multicolumn{2}{c}{Single star models} & \multicolumn{2}{c}{Binary models} & \multicolumn{2}{c}{Binary models w/ leaky photons} & & \\
\hline
 & Age (env)$_{\rm{sin}}$ & \Msin  & Age (env)$_{\rm{bin}}$ & \Mbin & Age (env)$_{\rm{corr}}$ & \Mcor & \Mlit & \Mlit \\
 & (Myr) & (M$_\odot$) & (Myr) & (M$_\odot$) & (Myr) & (M$_\odot$) & (M$_\odot$) & source \\
\hline\noalign{\smallskip}
SN~1999br & 8.4 & 23.8 & 25.1 & 10.4 & 10 & 19.9 & $<19.5$ & Pre-image$^1$ \\
SN~2001X	 & 10.1 & 19.8 & 39.8 & 8.0 & 12.6 & 17.3 & $13-15$ & Nebular phase$^2$ \\
SN~2004dg & 7.3 & 26.8 & 15.8 & 14.0 & 15.8 & 14.0 & $<11.6$ & Pre-image$^1$ \\
SN~2006my & 11.3 & 18.5 & 79.4 & 6.0 & 20 & 11.7 & $13.9^{+2.9}_{-3.0}$ & Pre-image$^1$ \\
SN~2008cn & 11.5 & 18.3 & 100 & 5.5 & 15.8 & 14.0 & $15.9^{+1.2}_{-1.1}$ & Pre-image$^1$ \\
SN~2009H & 8.1 & 24.4 & 15.8 & 14.0 & 10 & 19.9 & $<18$ & Pre-image$^3$ \\
SN~2009N & 8.8 & 22.8 & $100$ & $5.5$ & 15.8 & 14.0 & $<13$ & Pre-image$^3$ \\
SN~2009md & 11.6 & 18.2 & 100 & 5.5 & 12.6 & 17.3 & $8.0^{+1.9}_{-1.5}$ & Pre-image$^1$ \\
SN~2012P & 7.2 & 27.1 & $12.6-15.8$ & $14.0-17.3$ & 12.6 & 17.3& $\sim 15$ & Nebular phase$^4$ \\
SN~2012ec & 9.7 & 20.6 & 39.8 & 8.0 & 10 & 19.9 & $16.8^{+1.4}_{-1.3}$ & Pre-image$^1$\\
iPTF13bvn & 11.2 & 18.6 & 50.1 & 7.3 & 10 & 19.9 & $10-12$ & Pre-image$^{5}$ \\
\hline
\end{tabular}
\begin{tablenotes}
\small\item $^1$ \citet{db18}, $^2$ \citet{spw+17}, $^3$ \cite{sma15}, $^4$ \citet{fst+16}, $^5$ \citet{em16}
\end{tablenotes}
\end{threeparttable}
\end{table*}

\subsubsection{Binary models corrected for photon leakage}
\label{sssec:leakybinfits}
The results described above may be an effect of photon leakage, whereby a significant fraction of ionising photons produced by massive stars escape the circumstellar environment without interacting with the surrounding gas. This is likely to be an increasing problem in low density environments, such as is the case for the large majority of our SN sample, where electron densities are typically below $10^2$~cm$^{-3}$. By not taking into account this fraction of escaping photons, we under-estimate the UV emission from young stars, thus over-estimating the age of the stellar population at that position. \citet{xse18} tried to include the effect of leaky photons in their BPASS synthetic spectra, and found little change in their BPT emission line flux ratios, but a notable decrease in the model line equivalent width (in their case \hb) for the same given stellar population age. In \citet{xge+19} they revisited this issue, and used the difference between the observed and expected (by the models) \ha\ EW as a function of stellar population age to correct for the effect of photon leakage.

To test the effects of photon leakage, we use these corrections to convert our measured \ha\ EW to the \ha\ EW expected in the absence of any photon leakage. These corrections are available in the form of a look-up table produced by \citet{xse18}, and are dependent on stellar age and metallicity. We then re-did the analysis described above using the photon leakage-corrected \ha\ EW values to obtain new, best-matched stellar ages. These values, together with the best-matched ages where no photon leakage correction is applied, are given in Table~\ref{tab:SNage}. As expected, the stellar ages are greatly reduced when taking into account the effects of photon leakage, and the best-matched BPASS models now have ionised gas line ratios that are in much better agreement with our data. This is illustrated in Fig.~\ref{fig:leakyBPTs}, which is similar to Figs.~\ref{fig:allBPTs} and \ref{fig:SNBPTs}, but where the results for each SN are shown on separate rows, and in each case we only show the subset of binary BPASS models with the same $n_H$ and $Z$ as the model best-matched to the SN in question. The colour of the SN data points in Fig.~\ref{fig:leakyBPTs}, indicating its best-matched age, are now in general similar to the colour of the model curves that lie closest to the data points.

\subsubsection{Single star evolution models}
\label{sssec:sinfits}
For the single evolution models, we only use the observed EW(\ha) to constrain the most likely SPS models, firstly to be able to compare our results with  values reported in the literature, but also because the single star evolution models are in very poor agreement with the gas line emission ratios observed in \hii\ regions \citep[e.g.][]{xse18}. This is illustrated in Fig~\ref{fig:allBPTs}, where all the solid lines corresponding to the single stellar evolution models fall to the bottom left of the SDSS galaxies (grey scale) and SN positions, with only the youngest ($\la 3$~Myr), and more metal poor ($\la 0.4$Z$_\odot$) stellar populations lying close to observations. This poor fit between single star synthesis models and data on the BPT diagram is true of SPS models in general that do not consider the effects of binary interaction or stellar rotation in the evolution of stars \citep[e.g.][]{lkl10}.

In table~\ref{tab:SNage} we summarise the age that we derive for the stellar population at the SN positions based on our single and binary model analysis. Note that the binary evolution BPASS models that we compare our data with are in increments of 0.1~dex in logarithmic stellar age, such that the best-matched ages given in Table~\ref{tab:SNage} are limited to discrete values within the range 1~Myr to 100~Myr. In the case of the single stellar evolution models, because we only consider a single property (\ha\ EW), we can join the discrete stellar ages considered in the BPASS models on an \ha\ EW versus stellar age plot to produce a continuous distribution. The ages reported in Table~\ref{tab:SNage} for our analysis using the single star models are therefore not limited to discrete values. Any errors associated with our derived stellar ages are dominated by uncertainties in the BPASS models that we use rather than in our emission line measurements. Therefore, rather than report errors in Table~\ref{tab:SNage}, we discuss the limitations of our method in section~\ref{sec:disc}.

\subsection{Converting stellar age to \Minit}
\label{ssec:age}
In the absence of any mass transfer with a companion star, the evolutionary tracks followed by a star are primarily a function of \Minit\ and metallicity \citep[e.g.][]{hfw+03,gmw+09}. In this case more massive stars explode first, and it is therefore possible to work backwards, and infer the SN progenitor \Minit\ from a given stellar age and metallicity. This is not the case for binary systems, where mass transfer can have a significant effect on the stellar structure and evolution, removing any single correspondence between the \Minit, metallicity, and lifetime of a star \citep[e.g.][]{xge+19}. Nevertheless, although there are physical grounds to believe that binary population models should provide a more accurate description of the relation between stellar population age and the observed, circumstellar emission line spectra, that does not necessarily imply that the SN progenitors themselves are in binary systems.

There is evidence to show that SN type Ib originate from binary systems \citep{lgs+11,efs+13,kda+13a,fgv+16,gsm+16,lbj+16,kpm+17,kag+18,tsb+18,paj+19}, whereby the SN progenitor loses its hydrogen envelope through mass transfer to its companion star rather than through stellar winds. It is also thought that at least some SNe Ic are the result of binary star interactions, although single star progenitor channels may also exist. Normal SNe II, on the other hand, are thought to originate from progenitor stars that evolve effectively as single stars, without interacting with any binary companions. For example, pre- and post-explosion images have not revealed any compelling evidence for the presence of a binary companion in sufficiently close orbit to the SN to have influenced the progenitor’s evolution, and this applies to all SNe II in our sample that had progenitor detections in pre-explosion images. It is nevertheless possible that any prior companion star either merged with the SN progenitor, or it itself exploded, giving the surviving star (which we later see as a SN) a large kick, moving it away from its birth place \citep{zdj+19}. We shall discuss these possibilities in greater detail in section~\ref{sec:disc}, but for the purpose of being able to compare our results to the \Mlit\ values published elsewhere, we shall assume that our SN II effectively evolved as single stars. We shall refer to those stellar mass estimates inferred from direct SN or progenitor observations as \Mlit\ and masses from work based on the SN environment as \Mzams.

To estimate \Mzams\, we apply the latest single star isochrone models from \citet{mgl+17}, taking into account the metallicity and progenitor ages that we measured in the previous sections (Table~\ref{tab:SNage}). We use the notation \Msin\ and \Mbin\ to refer to initial mass estimates originating from single star or binary system evolution models, respectively, and \Mcor\ corresponds to binary system-based mass estimates where a correction for photon leakage has been applied. Although the model assumptions involved in converting our SN progenitor ages to a corresponding \Mzams\ add additional uncertainty to the progenitor properties that we derive, we make this step in order to be able to compare our results to the progenitor masses inferred from pre-explosion images and SN nebular spectral modelling; techniques that also assume that the SN progenitors evolved as single stars. There is one SN in our sample, iPTF13bvn, which was a type Ib, and for which there is compelling evidence that the SN progenitor had a companion star. In this case, our \Mzams\ estimate is therefore likely invalid, and we return to this point in section~\ref{sec:binarity}.

\subsection{\Minit\ compared to literature values}
\label{ssec:litcomp}
\subsubsection{Other \Mzams\ estimates based on \ha\ EW}
As mentioned in the introduction, the use of \ha\ EW to trace the SN progenitor age, and subsequently \Mzams, has been extensively used, most recently by \citet{kag+18}, which includes many of the SNe studied here (9/11). \citet{kag+18} considered only single star evolution models in their analysis, and used the Starburst99 SPS models \citep{lsg+99}. Differences between SPS models that assume single star evolution are relatively minor \citep{kga+16}, and indeed the largest inconsistencies between the results in \citet{kag+18} for the sample of overlapping SNe, and our single star model results originate from differences in the \ha\ EW measured at the SN position. These differences are likely related to the relatively large uncertainty in the SN position \citep[1$-$1.5 arcsec;][]{kag+18}, which we were able to reduce down to $<0.2$ arcsec for the SNe in our sample (section~\ref{ssec:SNpos}). Despite this, our initial mass estimates (from the single star models) are in fair agreement with the mass estimates from \citet{kag+18}, typically within 10$-$20\% of each other. The largest discrepancy is in the case of SN~2012ec, where \citet{kag+18} report a substantially larger \ha\ EW of $324.9\pm 36.3$\AA, compared to the $26.4\pm 0.1$\AA\ that we measure. It is not clear what the source of this large discrepancy is, but there could be several explanations, including aperture size and background subtraction. The IFU observations reported in \citet{kag+18} were taken with the VLT/Visible Multi-Object Spectrograph (VIMOS) rather than with MUSE, and the much smaller 13 arcsec $\times$ 13 arcsec FoV thus complicates background subtraction. 

Another promising technique that may allow us to constrain the ages of older progenitor systems such as SN II, but without the need for very sensitive or high spatial resolution data, is using the SN light curve itself \citep[e.g][]{lbj+16,exs+18}, which is directly related to progenitor and explosion properties. In a similar way to this paper, \citet{egr+19} applied their supernova light curve population synthesis models to constrain the progenitor initial mass of a sample of 11 SNe IIP with progenitor detections in pre-explosion images, 3 of which overlap with our sample: SN~2006my, SN~2009md and SN~2012ec. Their initial mass estimates for SN~2006my and SN~2012ec are consistent with our \Mcor\ estimates (although \Minit\ is largely unconstrained from the SN~2006my light curve). On the other hand, for SN~2009md we obtain a value for \Mcor\ that is far larger than the initial mass estimate in \citet{egr+19}, and instead our \Mbin\ is in closer agreement with the initial mass estimate of \citet{egr+19} and with \Mlit.

\begin{figure}
\includegraphics[width=1.1\linewidth]{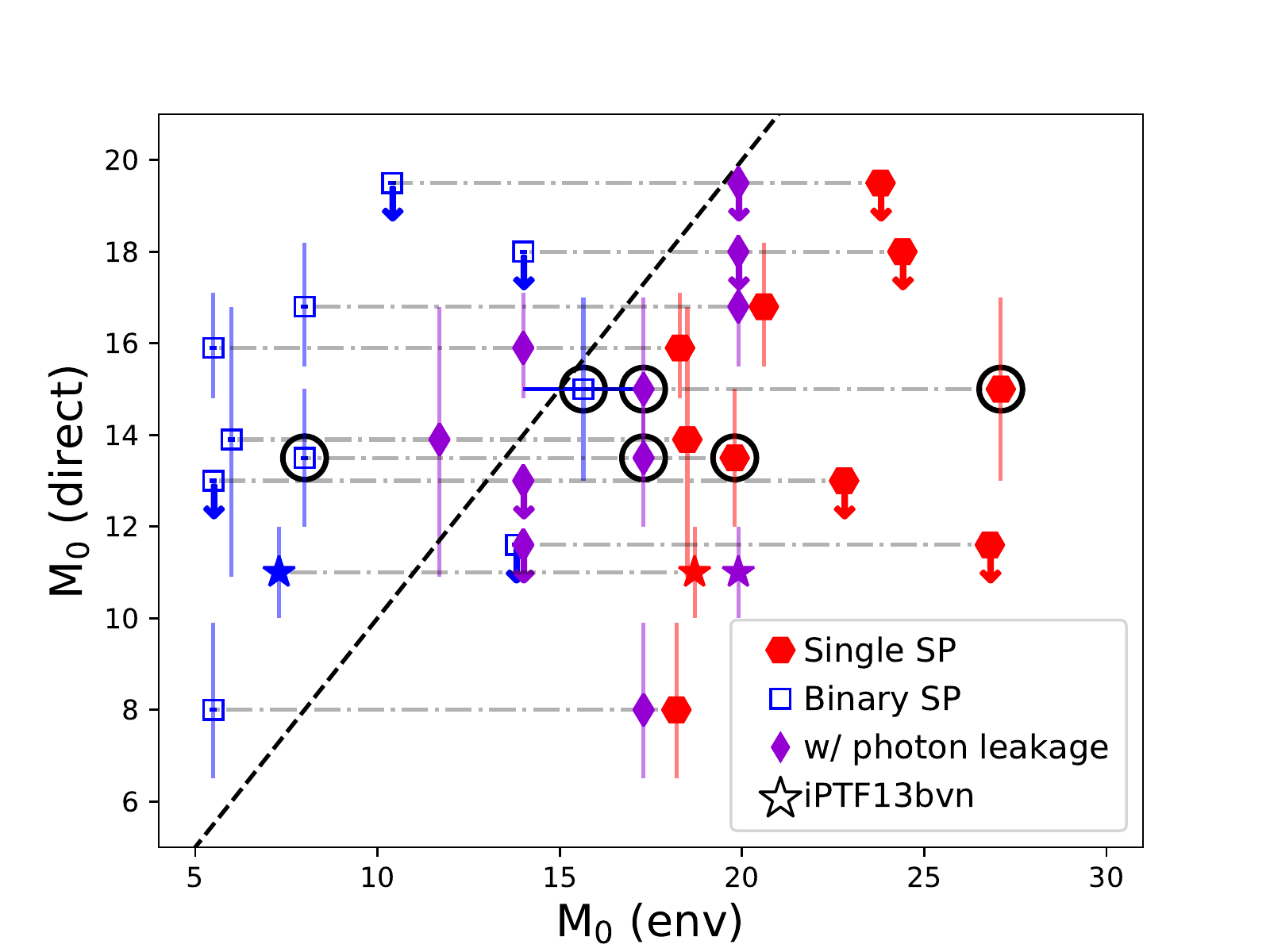}
\caption{\Mzams\ derived from our ionised gas analysis compared to previously reported \Mlit\ values in the literature from either pre-explosion images or from SN nebular spectral modelling (symbols outlined with black circles). The \Mzams\ values based on single stellar evolution models are plotted as solid red hexagons, open blue squares symbols correspond to the results from the binary system models, and masses where a correction has been made for photon leakage are plotted as solid purple diamonds. Grey, dot-dashed, horizontal lines connect data points corresponding to the same SN. \Mlit\ upper limits are shown with down-pointing arrows, and the star represents iPTF13bvn. The solid black line shows where \Mlit\ is equal to \Mzams. In the case of SN~2004dg, \Mbin = \Mcor, and we thus have slightly offset \Mbin\ to the left for clarity.}\label{fig:MzamsComp}
\end{figure}

\subsubsection{Our \Mzams\ versus \Mlit}
Nine of the SNe in our sample have published \Minit\ measurements derived from progenitor observations in pre-explosion images \citep{sec+09,sma15,db18}, and the remaining two have \Minit\ estimates based on SN nebular spectral modelling \citep{spw+17,fst+16}, all of which we refer to a \Mlit. These previously published progenitor masses are given in Table~\ref{tab:SNage}, as well as the corresponding references. In Fig~\ref{fig:MzamsComp} we compare \Mlit\ to \Msin\ (red hexagons), \Mbin\ (blue squares), and \Mcor\ (purple diamonds). Upper limits on \Mlit\ are indicated with a downward arrow, and to highlight iPTF13bvn in the figure, which was a type Ib SN and likely had a companion star, it is plotted with as a star.

From this figure it is immediately clear that results based on comparing the measured \ha\ EW to single star evolution models overpredict \Minit\ by up to a factor of two, corresponding to an age difference of 10$-$30~Myr. Conversely, comparing \ha\ EW and the BPT emission line ratios to binary stellar evolution models predominantly underestimates \Minit. This is consistent with the findings presented in \citet{xse18}, where they found that progenitor ages derived from binary evolution models for a sample of type II SNe were far greater than implied by the distribution in \Mlit\ determined from progenitor detections in pre-explosion images \citep{sma09,sma15}. If, similarly to \citet{xse18}, we make a correction for the escape of ionising photons that do not interact with the gas in the vicinity of the SNe (and as such to not contribute to the strength of the ionised gas emission lines), then we do find much better agreement between our initial mass estimates and \Mlit. There does, nevertheless, remain significant scatter between \Mcor\ and \Mlit. In part this may be due to inaccuracies in \Mlit, which are also model dependent, but it is also likely related to the large uncertainty in the fraction of the emitted, ionising photons that escape the star forming region. Despite this uncertainty, in low density environments such as those found in the vicinity of our SN sample, we would expect a notable fraction of ionising photons to escape the star forming region without undergoing any interactions. Our \Mbin\ values can thus be considered to be lower limits.

Of note is the fact that five of the SNe in our sample have \Mbin$<8$M$_\odot$, which is too low mass for a single star progenitor to have undergone core-collapse. This may imply that the star evolved in a binary system, where it is possible for stars with \Minit$\sim 5$~M$_\odot$ to merge or accrete sufficient mass to get above the core-collapse limit of 8~M$_\odot$ \citep[e.g.][]{zdI+17,esx+17}. Indeed, one of the SNe with \Mbin$<8$~M$_\odot$ is iPTF13bvn. Such a scenario would invalidate our derivation of \Mzams. An alternative possibility is that the progenitor stars moved away from their birth cloud. We explore these possibilities further in section~\ref{sec:disc}.

\subsection{Other environmental estimates of M$_\mathrm{ZAMS}$}
Given the uncertainty in any set of stellar population models on the fraction of stars in binary or multi-body systems, it is interesting to consider how this limitation of the models may affect other environmental studies that aim to constrain the SN progenitor. As mentioned in the introduction, the colours and magnitudes of resolved stars in the vicinity of the SN explosion have also been used to constrain the age of the SN progenitor \citep[e.g][]{mjw+11,jwm+14,mau17,dmr+18,wpm+14,whm+18,alb+19}. This technique relies on SPS models that produce isochrones for different ages on the colour-magnitude diagram (CMD). The position of these isochrones will vary depending on the fraction of binary systems and on the mass ratio distribution of the stars in binaries. Binary evolution causes stars to remain bluer and hotter for longer \citep{es09}, effectively moving their position on the CMD diagram up and to the left relative to single stellar evolution tracks of the same given age. In the presence of a significant fraction of interacting binary systems, the result of only considering single stellar evolution models will thus be to underestimate the age of the stellar population, especially for star clusters that are older than a few Myr. 

Considering only single star evolution models, \citet{whm+18} found that for a sample of 8 SN II with pre-explosion progenitor detections, their CMD-determined \Mzams\ estimates were consistent with \Mlit, contrary to our results. However, the associated uncertainties on their \Mzams\ were generally quite large ($> 40$\%), which may dominate over the uncertainties associated with the CMD isochrones used. 

In order to have a better understanding on the origin of differences in \Minit\ from direct progenitor and SN modelling, from our analysis presented here, and from other environmental-based studies, in the next section, we discuss the principle steps and assumptions involved in our analysis, and how they may introduce error in our \Mzams\ estimates.

\section{Discussion}
\label{sec:disc}
The principle assumptions involved in our analysis can be divided into three principle steps. The first is our assumed association between the emission line properties that we measure and the stellar population that hosted the SN progenitor. In making this association, we are assuming that the underlying emission at the position or in the vicinity of the SN is produced predominantly by a single stellar population, and that the SN progenitor was a member of this stellar population. The second is in the reliability of the SPS models used in our analysis, which are very sensitive to whether stars are treated in isolation or as interacting. Finally, to infer an \Mzams\ from the measured stellar population age, we assume that the SN evolves as a single star. If this is not the case then from these data alone it is not possible to translate an age to a progenitor \Minit. Below we discuss the validity and implications of each of our assumptions in detail.

\subsection{Star formation history and chance alignment}
\label{sec:sfh}
For typical dispersion velocities of a few km~s$^{-1}$ \citep{bg06}, the stars within a stellar cluster of age $\sim 30$~Myr will have dispersed over a region 100$-$200~pc across. However, stellar associations much older than this become increasingly diffuse as member stars migrate further apart. The effect is to decrease the stellar and ionised gas emission measured at the SN location, thus reducing the sensitivity of our observations, but more importantly, it also increases the probability of a chance alignment with an unassociated star forming region. 

Chance alignment of an unrelated \hii\ region will only affect our results if the unassociated \hii\ region is significantly younger (and hence brighter) than the stellar population that hosted the SN progenitor, which in turn would cause the SN progenitor age to be underpredicted. The \Mbin\ estimates that we derive are generally in agreement, or smaller than \Mlit, implying that chance alignment is not a cause for concern. However in the case of the single star BPASS models, the finding that \Msin\ is typically larger than \Mlit\ (Fig~\ref{fig:MzamsComp}) may imply, in at least some cases, that there is a chance alignment with a younger, unrelated star forming region coincident with the SN position.

The projected surface area of the SN host galaxies covered by the MUSE field of view with \ha\ EW $>20$\AA\ (corresponding to star population ages older than $\sim 10$~Myr in single stellar models) ranges from $\sim 10$\% to $\sim 50$\%, which provides an indication of the probability of an unrelated, young \hii\ region being projected onto the SN position. However, the probability of chance alignment will vary as a function of the radial position of the SN from the galaxy centre. To try and take this into account, we therefore consider the fraction of pixels with an \ha\ EW$>20$\AA\ within a 200 pc wide elliptical annulus centred at the galaxy nucleus and covering the SN location. From this analysis we find that, on average, 35\% of the area considered has an \ha\ EW$>20$\AA, and the median value is 28\%. From a sample of eleven SNe, we can thus expect around 3 SNe in our sample to lie in a region of their host galaxy that is coincident with a brighter but unrelated star forming region. Therefore, although there may be some cases where the equivalent widths that we measure may in fact be dominated by emission from an \hii\ region that is unrelated to the SN (e.g. see Fig.~\ref{fig:env}, right panel), it seems unlikely that this effect alone is the cause for the differences that we see between \Mlit\ and the \Msin, where the whole sample has \Msin$>$\Mlit.

\subsection{Runaway star}
\label{ssec:runaway}
Another possibility to consider is whether the SN progenitors could have migrated away from their natal star forming region (Fig~\ref{fig:env}). First we consider only those SNe with \Mbin$<8$M$_\odot$. Isolated stars with an initial mass \Minit$\sim 8$M$_\odot$ are expected to collapse to form a SN after $\sim 40$~Myr, which in the binary stellar evolution BPASS models, corresponds to the age of a stellar population that would produce an \ha\ EW=20$-$30\AA\ (e.g. SN~2008cn and SN~2012ec). For those SNe where we measure progenitor masses \Mbin$<8$M$_\odot$, we thus use this as a rough guide to find the nearest stellar population to the SN that has an age of $\sim 40$~Myr (according to binary stellar evolution models). With the exception of iPTF13bvn, which is very likely to have had a binary companion star, we estimate that the other SNe with very low \Mbin\ must have travelled from $\sim 100$~pc to $\sim 650$~pc (in the case of SN~2008cn) in order to have originated from a stellar population that is sufficiently young to still contain isolated stars that can undergo core-collapse. Due to projection effects, these distances are in fact effective lower limits, and imply that the progenitor stars moved away from their birth clouds at a minimum speed of 2$-$15~km/s for most of their lifetime.

Stellar velocities of a few km/s are often observed \citep[e.g.]{bg06}, and faster speeds of 10$-$15~km/s can be achieved if the star receives a kick from the SN explosion of a companion star \citep[e.g.][]{rzd+19}. \citet{elt+11} used binary population synthesis modelling to investigate the probable range in velocities expected from progenitors of varying SN types, and thus the typical distances that may exist between SN birth clouds and where they eventually explode. Their models imply that the progenitors of $\sim 25$~\% of SN IIP are the secondary stars of a binary system, and can receive typical kick velocities of $17.5\pm 18.7$~km/s when their primary star explodes. They found a similar range in velocities for type Ib SNe with secondary star progenitors. This is fully consistent with the velocities calculated above that we would need for our SN sample to be associated with star forming regions with EW$>20$\AA.

Nevertheless, in order to bring \Mbin\ in agreement with \Mlit\ (rather than just $>8$~M$_\odot$), the progenitors of several SNe in our sample would need to travel larger distances than discussed above, and thus be moving with velocities $> 15$~km/s. For example, to associate SN~2008cn to a star forming region sufficiently young to still contain stars with \Minit$\sim 16$~M$_\odot$ (the corresponding \Mlit\ value), the progenitor would have had to travel over 2.5~kpc from its birth cloud, corresponding to a velocity of $\sim 150$~km/s. If we carry out a similar analysis for SN~2001X and SN~2012ec, we find that the nearest stellar populations that are still likely to contain singley evolving stars with an \Mbin\ equal to the corresponding \Mlit\ are at projected distances of $>450$~pc and $>700$~pc, respectively, implying progenitor star velocities of $>25$~km/s and $>50$~km/s. Although velocities of up to $\sim 40$~km/s for SN IIP progenitors are within the range of expected values calculated in \citet{elt+11}, larger velocities, as in the case of SN~2008cn and SN~2012ec seem quite unlikely. This is especially true when we consider that `only' 25~\% of SN IIP are expected to receive a significant kick velocity from a binary companion according to the analysis of \citet{elt+11}. We therefore investigate other, more likely explanations for the discrepancy between \Mzams\ and \Mlit.

\subsection{Stellar population synthetic spectra}
\label{ssec:nebspec}
There are several model assumptions that go into relating observed gas emission line properties at the SN position to a stellar population age, in particular in the case of population models that include binary systems. Binary systems are complex, and their evolution depends greatly on a number of parameters, such as the mass ratio, and initial orbital period. In the BPASS v2.1 models, the distribution in mass ratio and initial period are assumed to be flat, and a cap of $10~000$ days is put on the period distribution in order to confine the fraction of massive stars in interacting binaries to 80\%, which is comparable to observations in the Milky Way \citep[e.g.][]{sdd+12,md17}. However, this may be too simplistic, and it may be that the fraction of massive stars in binaries changes as a function of environment \citep[e.g.][]{dl18}. The latest BPASS models (v2.2.1) contain a small population of single stars in additional to binary systems, and in the future it would thus be interesting to compare the ionised gas synthetic spectra produced from these models to observations. For now, however, some insight can be gauged on what the difference between the v2.1 and v2.2 model fits would be, since changing the stellar binary fraction should have a similar effect to correcting for photon leakage, which we investigated in \ref{sssec:leakybinfits}.

A more fundamental issue may be in the limitation of using emission lines to trace stellar properties in low density environments. At the position of our SN sample, we find that the electron densities are typically below $10^2$~cm$^{-3}$, and thus a significant fraction of ionising photons produced by massive stars may escape the circumstellar environment without interacting with the surrounding gas, resulting in `leaky photons'. To try and take this account we used corrections derived in \citet{xse18} to account for the observed decrease in the \ha\ EW as a result of photon leakage. This leaky photon model can also be thought of as a simple way of varying the initial binary fraction, since decreasing the binary fraction would similarly decrease the model stellar population age for the same flux of ionising photons.

The results are shown in Fig.~\ref{fig:MzamsComp}, and show that the correction goes too far in that for most SNe, \Mcor$>$\Mlit. SPS models with a smaller stellar binary fraction (e.g. BPASS v2.2) that also take into account the effects of photon leakage will thus likely make the discrepancy between \Mcor\ and \Mlit\ even larger. Nevertheless, the precise correction that should be applied to account for leaky photons is hard to estimate, and is likely a function of several properties, such as electron density, stellar population age, and ionisation parameter. As a general note, the environmental conditions at the position of our SN sample are not representative of typical \hii\ regions, and the assumptions on the conditions of the surrounding gas that go into producing the synthetic spectra used in our analysis are thus very uncertain. The small electron densities that we measure at the location of our SNe may well imply that photon leakage is a prime cause for the large discrepancy that we find between \Mbin\ (taking no account of leaky photons) and \Mlit, and more detailed analysis on the extent of this effect using radiative transfer codes such as {\sc cloudy} may thus be warranted.

Other important uncertainties in the model assumptions are the input IMF, for example, although again the effects should be somewhat degenerate with changing the stellar binary fraction and with making a correction for photon leakage. By considering single star and v2.1 binary star (without photon leakage corrections) BPASS models in our analysis, we are effectively straddling the two extremes in the predicted stellar ages. Greater certainty on what the assumed IMF and binary fraction should be, for example,  will require greater observational constraints to guide future stellar population models.

\subsection{The importance of companion stars in SN explosions}
\label{sec:binarity}
The final, but an important assumption in our analysis in converting stellar age to an \Mzams\ is that the progenitors in our SN sample evolved as single stars. As mentioned in section~\ref{ssec:age}, iPTF13bvn was a SN Ib, which are largely believed to have undergone a period of mass exchange prior to explosion \citep[e.g][]{efs+13}. The progenitor of iPTF13bvn was detected in pre-explosion images \citep{cka+13}, and the colours of the progenitor were consistent with the star having evolved in a binary system \citep{bbf+14,efm+15,em16,fgv+16,fst+16}. This would make our assumptions in deriving an \Mzams\ invalid, which is the reason why we highlight this SN in Fig.~\ref{fig:MzamsComp}.

Partially stripped SN type IIb may have similar progenitors to SN Ib, and there is indeed observational evidence that at least some SN IIb progenitors underwent a mass accretion phase: e.g. SN~1993J \citep{msk+04}, SN~2008ax \citep{ces+08}, SN~2011dh \citep{agy+11,vlc+11}. There is one SN in our sample, SN2012P, that is of type IIb \citep{agb+12}, and although there is no observational claim that this particular SN progenitor evolved within an interacting binary, it is of course possible. 

The rest of our sample were `standard' SN II (i.e. SN IIP or SN IIL), and where relevant, none had a binary companion star detected in pre- or post-explosion images, although this again does not rule out a mass accretion phase. \citet{zdj+19} found from their population synthesis simulations that between 1/3 and 1/2 of SN II progenitors may have interacted with a binary companion prior to explosion, as was also found by \citet{elt+11} using a simpler analysis to determine the SN type. We may expect those SN II with progenitors in interacting binaries to produce more exotic types of hydrogen-rich SNe than are present in our sample. For example, some type IIn or anomalous SN~1987A-like events have shown evidence of mass accretion episodes prior to explosion \citep[e.g.][]{st15,hm89,pjr90,tsr+13,muh19}. However, based on SN rates \citep[e.g.][]{sec+09}, this does not account for 1/3$-$1/2 of SN II population.

Importantly, \citet{zdj+19} find that most SNe in binary systems should not necessarily have a detectable companion star in spatially resolved observations of the SN field. From their simulations, most SNe II with progenitors that interacted with a binary companion star either merged with the companion star prior to explosion, or were ejected from the binary system due to the prior explosion of the companion star. These results would imply that 3$-$5 of the 9 SNe type II in our sample evolved within binary systems, and thus in these cases, \Minit\ estimates based either on environmental studies or pre-explosion progenitor detections would be inaccurate due to incorrect model assumptions.

Despite these uncertainties on the effect of binary interactions on the progenitors of our SN sample, the progenitor age estimates presented in this paper should be more reliable. Whereas spectra of star forming regions are sensitive to stellar age (albeit model dependent), a star that has accreted material from a binary companion through mass transfer may have a very different `effective initial mass'.

\section{Summary and conclusions}
\label{sec:concl}
The current extensive availability of sensitive IFU data of nearby galaxies opens the possibility for very detailed analysis of the nearby ($<100$~pc) environmental properties of large samples of SNe. For example, the All-weather MUse Supernova Integral field Nearby Galaxies (AMUSING) survey \citep[MUSE;][]{gar+16} has now been running for 4 years, and currently contains a sample of $\sim 400$ SN host galaxies, $\sim 30$\% with a physical spatial resolution of $\sim 200$~pc or better. Driven by this opportunity, in this paper we have investigated the reliability of using the \ha\ EW and ionised gas line ratios at SN positions to constrain the SN progenitor age, and thus its zero-age main-sequence mass, \Minit. If validated, such a tracer would offer the opportunity to greatly increase the sample of SNe with \Minit\ estimates.

We selected a sample of SNe with high spatial resolution MUSE host galaxy observations (better than 150~pc), and with reliable \Minit\ estimates from progenitor detections in pre-explosion images, or from SN nebular phase spectral modelling. This latter criterion was essential to be able to verify mass estimates in this work that are based on the SN nearby environmental properties, \Mzams. We compared the \ha\ EW and key diagnostic ionised gas line ratios at the SN positions to synthetic spectra of ionised gas produced with the BPASS stellar evolution models, assuming both single and binary stellar evolution. The synthetic spectra in closest agreement with our observations provided us with an estimate of the age of the stellar population at the SN location, from which we derived the progenitor star initial mass, \Minit.

Our principle result is that when comparing \Mzams\ to direct mass estimates, \Mlit, the BPASS single star models systematically overestimate \Minit\ by 
typically $>5$~M$_\odot$, whereas the binary stellar population models tend to significantly underestimate \Minit. We explore possible reasons for the differences between \Mzams\ and \Mlit, and conclude that at least in the case of older progenitors, such as those of type II SNe, single star stellar populations are not valid. This conclusion is all the more strengthened when one considers that there should be a notable amount of photon leakage in the low-density environments in which our SN sample reside, which would further increase \Msin.

Runaway stars may partly account for the poor agreement between \Mbin\ and \Mlit, whereby up to 25\% of our SNe may have exploded a few hundreds of parsec away from their natal region \citep{elt+11}. However, we find that a more important issue is likely to be that a significant fraction of the ionising photons produced by massive stars escape without ionising the surrounding gas. This effectively weakens the relation between the ionising radiation from massive stars and the emission line properties from the excited gas, thus limiting the use of emission lines as tracers of stellar population age to only the youngest star forming regions. Further, detailed analysis to try and model more accurately the low-density gas typically found at the position of SN II would be desirable in order to produce more reliable synthetic spectra of the environments of SN II.

Another caveat in our analysis is that, in inferring an initial mass from a progenitor stellar age, we assume that the SN progenitor star evolved effectively as an isolated star, which may in fact not be the case \citep[e.g][]{elt+11, zdj+19}. Thus, although the progenitor ages that we measure would be unaffected by this assumption, the true initial mass, \Minit, may be very different to the values we estimate. This is of course also true for other methods of measuring \Minit. If a significant fraction of SN II progenitors are found to have interacted with a binary companion, then initial mass estimates based on age-dating nearby stars or on progenitor pre-explosion detections, for example, may also need revising. 

This is not to say that the nearby environments of SNe cannot help us constrain their progenitor properties. Environmental studies continue to be important for determining the relative differences between the progenitors of varying SN type \citep[e.g.][]{gar+16,gsm+16,gas+18,kag+18}. Furthermore, in the case of very young stellar explosions, such as gamma-ray bursts or probably type Ic SNe, the parent stellar population and gas in the natal \hii\ region has had little time to diffuse away, thus reducing the effects of photon leakage, and increasing the sensitivity of emission lines as tracers of stellar age. This is supported by the good agreement in the ages inferred from the \ha\ EW for a sample of young stellar clusters of known age \citep{kga+16,kks+17}.

For older progenitor stars, such as SN II, analysis of the stellar Balmer absorption lines on spectra of the nearby SN environments may offer a promising alternative of tracing stellar age. Stellar absorption lines have the advantage that they become increasingly pronounced for older stellar populations, in contrast to the \ha\ EW line, although to be reliable, spectra should cover the Balmer series shortward of \hb. With MUSE this is only possible for sources that are at redshifts $\ga 1.2$, but then the spatial resolution is limited to kpc scales. Bluer, albeit lower spatial resolution IFU instruments are thus better suited for this purpose, such as with the Potsdam Multi Aperture Spectograph (PMAS), as was done by \citet{gas+18}. Nevertheless, to be able to judge how reliable any indirect method is of tracing the progenitor stellar age or initial mass, it is important to verify results, using a sample of SNe with \Minit\ estimates based on more direct, and thus reliable techniques.

\section*{Acknowledgements}
This study is based on data acquired at ESO, Programme ID 095.D-0172(A), 095.B-0532(A), 097.B-0165(A), and 099.B-0242(A). We would like to thank the anonymous referee for their constructive feedback, which helped improve the quality of the paper. P.S. would like to acknowledge the work and help from Dr Thomas Kr{\"u}hler in developing the tools used in this paper to analyse the MUSE data presented in this work. This work made use of v2.1 of the Binary Population and Spectral Synthesis (BPASS) models as last described in \citep{esx+17}. P.S. acknowledges support through the Sofja Kovalevskaja Award from the Alexander von Humboldt Foundation of Germany. T.-W. C. acknowledgments the funding provided by the Alexander von Humboldt Foundation. L.G. was funded by the European Union's Horizon 2020 research and innovation programme under the Marie Sk\l{}odowska-Curie grant agreement No. 839090.

\bibliography{refs}

\appendix
\section{Ionised gas emission lines and oxygen abundances}
Given the large variety of metallicity diagnostics that are available, which can vary by up to 0.7~dex \citep{ke08}, in Table~\ref{tab:Zall} we provide the metallicity that we measure at the positions of our SNe sample for five commonly used metallicity diagnostics that can be applied to the MUSE data. These are the \citet{dks+16} N2S2 diagnostic, the \citet{pp04} O3N2 and N2 diagnostics, referred to as PP04 O3N2 and PP04 N2 in Table~\ref{tab:Zall}, and the \citet{mrs+13} re-calibration of the O3N2 and N2 diagnostics (M13 O3N2 and M13 N2 in Table~\ref{tab:Zall}). In Table~\ref{tab:emlines} we also provide the main strong, ionised gas emission lines measured at the SN positions, which were used to compute the gas-phase metallicity, and which are needed to place the SN environments on the three, diagnostic BPT diagrams discussed in the paper.

\begin{table*}
\caption{Host galaxy gas phase metallicities measured at the location or in the vicinity of the ccSNe in our sample using a number of metallicity diagnostics\label{tab:Zall}}
\centering
\begin{threeparttable}
\begin{tabular}{l|ccccc}
\hline
\hline\noalign{\smallskip}
{SN} & \multicolumn{5}{c}{12+log(O/H)} \\
 & N2S2$^{\rm a}$ & PP04 O3N2 & PP04 N2 & M13 O3N2 & M13 N2 \\
\hline\noalign{\smallskip}
SN~1999br & $8.65\pm 0.08$ & $8.54\pm 0.11$ & $8.58\pm 0.02$  & $8.41\pm 0.11$ & $8.49\pm 0.02$ \\
SN~2001X  & $9.01\pm 0.05$ & $8.83\pm 0.30$ & $8.64\pm 0.01$  & $8.60\pm 0.29$ & $8.51\pm 0.01$ \\
SN~2004dg & $-$ & $8.87\pm 0.21$ & $8.65\pm 0.01$  & $8.63\pm 0.20$ & $8.52\pm 0.01$ \\
SN~2006my & $8.50\pm 0.17$ & $8.71\pm 0.21$ & $8.69\pm 0.04$  & $8.52\pm 0.21$ & $8.54\pm 0.04$ \\
SN~2008cn$^{\rm b}$ & $-$ & $8.66\pm 0.36$ & $8.73\pm 0.04$  & $8.49\pm 0.35$ & $8.55\pm 0.04$ \\
SN~2009H & $-$ & $8.71\pm 0.10$ & $8.64\pm 0.00$  & $8.52\pm 0.10$ & $8.52\pm 0.00$ \\
SN~2009N$^{\rm b}$ & $8.62\pm 0.55$ & $8.60\pm 0.15$ & $8.40\pm 0.03$  & $8.45\pm 0.15$ & $8.38\pm 0.03$ \\
SN~2009md & $8.49\pm 0.28$ & $8.60\pm 0.23$ & $8.62\pm 0.05$  & $8.44\pm 0.22$ & $8.51\pm 0.05$ \\
SN~2012P & $-$ & $8.88\pm 0.05$ & $8.60\pm 0.00$  & $8.64\pm 0.05$ & $8.50\pm 0.00$ \\
SN~2012ec & $-$ & $8.67\pm 0.09$ & $8.64\pm 0.01$  & $8.50\pm 0.09$ & $8.52\pm 0.01$ \\
iPTF13bvn & $9.03\pm 0.14$ & $8.83\pm 0.36$ & $8.69\pm 0.02$  & $8.60\pm 0.35$ & $8.54\pm 0.02$ \\
\hline
\end{tabular}
\begin{tablenotes}
\small\item $^{\rm a}$ The \citet{dks+16} metallicity diagnostic is only valid for gas phase metallicities $12+\log\rm{(O/H)}\la 9.05$
\small\item $^{\rm b}$ Analysis done in region nearby to SN position which has \ha\ EW $>10$ \AA. 
\end{tablenotes}
\end{threeparttable}
\end{table*}

\begin{table*}
\caption{Ionised emission line properties at the location of the 11 ccSNe in our sample \label{tab:emlines}}
\centering
\begin{threeparttable}
\begin{tabular}{l|ccccccc}
\hline
\hline
 & \multicolumn{6}{c}{Emission line flux ($10^{17}$ erg/s/cm$^2$)} & E(B$-$V)$_{\rm host}$ \\
SN & \ha & \hb & [\ion{O}{iii}] & [\ion{O}{i}] & [\ion{N}{ii}]$\lambda$6583 & [\ion{S}{ii}]$\lambda$6730 & \\
\hline
SN~1999br   &  $ 15.02\pm 0.01$ & $  4.39\pm 0.04$ & $ 4.86\pm 0.04$ & $ 0.82\pm 0.11$ & $ 4.22\pm 0.01$ & $  6.32\pm 0.06$ & 0.14 \\
SN~2001X	   & $101.19\pm 0.04$ & $ 27.80\pm 0.11$ & $ 4.43\pm 0.15$ & $ 2.14\pm 0.17$ & $32.55\pm 0.04$ & $ 26.01\pm 0.13$ & 0.18 \\
SN~2004dg   &  $ 43.06\pm 0.01$ & $  9.90\pm 0.03$ & $ 1.28\pm 0.03$ & $ 0.58\pm 0.05$ & $14.44\pm 0.01$ & $ 11.94\pm 0.08$ & 0.32 \\
SN~2006my   &  $ 13.00\pm 0.02$ & $  4.44\pm 0.05$ & $ 1.87\pm 0.04$ & $ 0.71\pm 0.10$ & $ 4.67\pm 0.02$ & $  7.02\pm 0.14$ & 0.00 \\
SN~2008cn   &  $  6.25\pm 0.01$ & $  1.71\pm 0.05$ & $ 0.23\pm 0.22$ & $ 0.34\pm 0.17$ & $ 2.38\pm 0.01$ & $  1.73\pm 0.20$ & 0.09 \\
SN~2008cn$^{\rm a}$ & $ 10.79\pm 0.01$ & $  2.89\pm 0.05$ & $ 1.90\pm 0.07$ & $ 0.43\pm 0.11$ & $ 4.22\pm 0.02$ & $  2.69\pm 0.26$ & 0.13 \\
SN~2009H	  &   $ 56.39\pm 0.01$ & $  9.82\pm 0.04$ & $ 3.82\pm 0.04$ & $ 2.02\pm 0.02$ & $18.25\pm 0.01$ & $ 19.15\pm 0.02$ & 0.58 \\
SN~2009N    &  $  2.69\pm 0.34$ & $  0.50\pm 0.54$ & $ 0.30\pm 0.39$ & $-$ & $ 0.03\pm 0.34$ & $  2.02\pm 0.24$ & 0.53 \\
SN~2009N$^{\rm a}$  & $ 16.98\pm 0.01$ & $  4.47\pm 0.03$ & $ 1.91\pm 0.03$ & $-$ & $ 2.78\pm 0.01$ & $  5.05\pm 0.32$ & 0.23 \\
SN~2009md   &  $  5.55\pm 0.01$ & $  1.64\pm 0.04$ & $ 1.35\pm 0.01$ & $ 1.44\pm 0.06$ & $ 1.71\pm 0.01$ & $  3.72\pm 0.12$ & 0.12 \\
SN~2012P	 &    $712.48\pm 0.03$ & $146.17\pm 0.09$ & $15.00\pm 0.09$ & $25.19\pm 0.20$ & $11.65\pm 0.03$ & $147.25\pm 0.60$ & 0.42 \\
SN~2012ec  &   $ 76.56\pm 0.02$ & $ 14.52\pm 0.06$ & $ 7.46\pm 0.07$ & $ 3.07\pm 0.03$ & $24.92\pm 0.02$ & $ 29.45\pm 0.03$ & 0.50 \\
iPTF13bvn  &  $ 16.29\pm 0.01$ & $  4.34\pm 0.04$ & $ 0.76\pm 0.03$ & $ 0.38\pm 0.13$ & $ 5.90\pm 0.01$ & $  5.03\pm 0.08$ & 0.20 \\
\hline
\end{tabular}
\begin{tablenotes}
\small\item $^{\rm a}$ Analysis done in region nearby to SN position which has \ha\ EW $>10$ \AA. 
\end{tablenotes}
\end{threeparttable}
\end{table*}
\newpage

\section{BPT properties and SN location and best-matched stellar ages}
\begin{figure*}
  \includegraphics[width=1.0\linewidth]{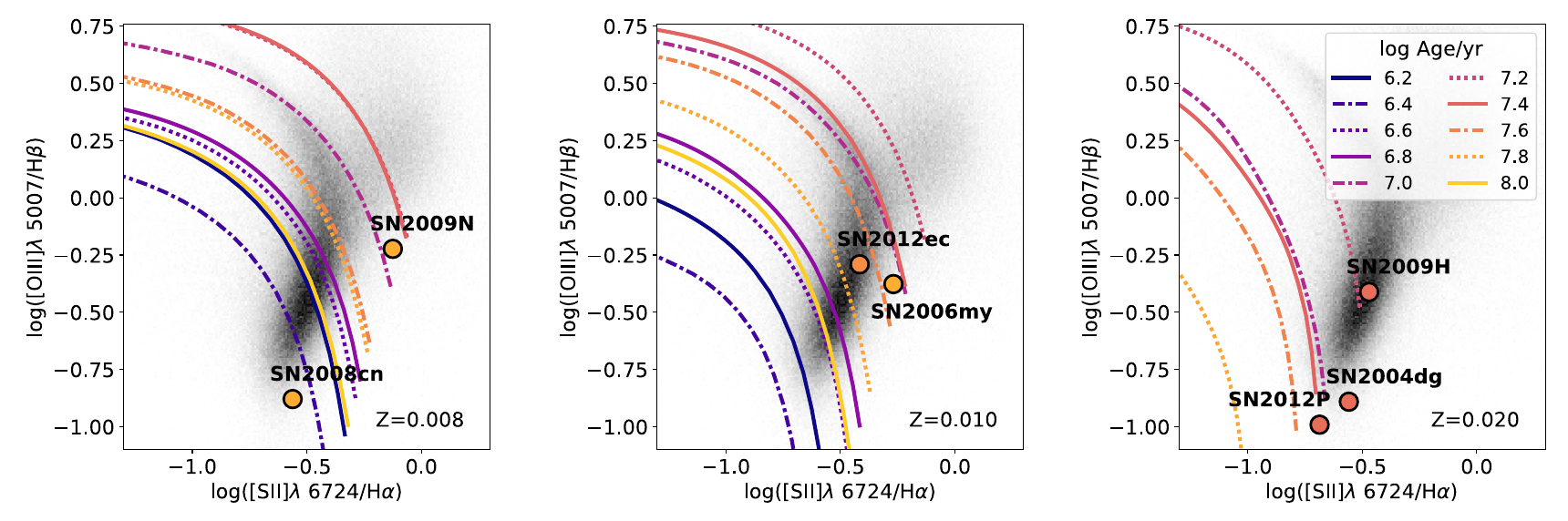}
\caption{BPT diagram of [\ion{O}{iii}]/\hb\ against [\ion{S}{ii}]/\ha, each with a different set of BPASS binary evolution models, colour-coded by stellar age. All curves plotted correspond to models with gas densities $\log$($\rm{n_H}$)=1.5, but they vary in metallicity, going from Z=0.008 (left), Z=0.010 (middle), to Z=0.020 (right). The SNe in each panel are plotted alongside binary evolution models of the same gas density and metallicity as their corresponding best-matched model, and the data points are colour-coded according to the best-matched stellar age (without applying correction for photon leakage) using the same key as for the model curves (indicated in the right hand panel).}\label{fig:SNBPTs}
\end{figure*}

To illustrate the discrepancy in the gas emission line ratios measured at the SN position and predicted by the best-matched binary evolution BPASS models, in Fig.~\ref{fig:SNBPTs} we plot a subset of the binary evolution models (coloured curves) together with some of the SNe (coloured data points) in our sample on the $\log$[\ion{O}{iii}]$\lambda$5007/\hb\ against $\log$[\ion{S}{ii}]$\lambda$6724/\ha\ BPT diagram, both colour-coded by stellar age. The model curves shown in each panel of Fig.~\ref{fig:SNBPTs} correspond to models with density $\log$($\rm{n_H}$)=1.5, but with different metallicities, which from left to right are Z=0.008, 0.010, 0.020. In each panel, the model curve of the same colour (i.e. stellar age) as any given SN data point corresponds to the model that is found to be best-matched to our observations according to the analyses described in section~\ref{sssec:binfits}. The fact that the colour of the SN data points (i.e. best-matched stellar age) are often of a different colour (i.e. stellar age) to the model curves that lie nearest to the SN data points (SN~2008cn in the left-hand panel is an extreme example) illustrates that the best-matched models are often not very representative of the conditions at the SN positions, and are being largely driven by the observed \ha\ EW. 

Fig.~\ref{fig:leakyBPTs} is similar to Fig.~\ref{fig:SNBPTs}, but we now show each SN separately (one row per SN), on each of the three diagnostic BPT diagrams (one BPT plot per column). For each row, the gas density and metallicity of the plotted set of binary evolution BPASS models is indicated in the left had panel, together with the name of the corresponding SN. The colour key is shown in the right-hand panel, and the colour of the SN data points now correspond to the best-matched age after applying a correction for photon leakage. By applying a correction to the measured \ha EWs to take into account photon leakage, the best-matched stellar population ages are generally younger than when no correction is applied, and the colour of the data points shown in Fig.~\ref{fig:leakyBPTs} (i.e. best-matched SN progenitor age) are now generally closer to the model curve of the same corresponding colour (i.e. stellar age) than was the case prior to the photon leakage correction (e.g. see Fig.~\ref{fig:SNBPTs}).

\begin{figure*}
  \includegraphics[width=1.0\linewidth]{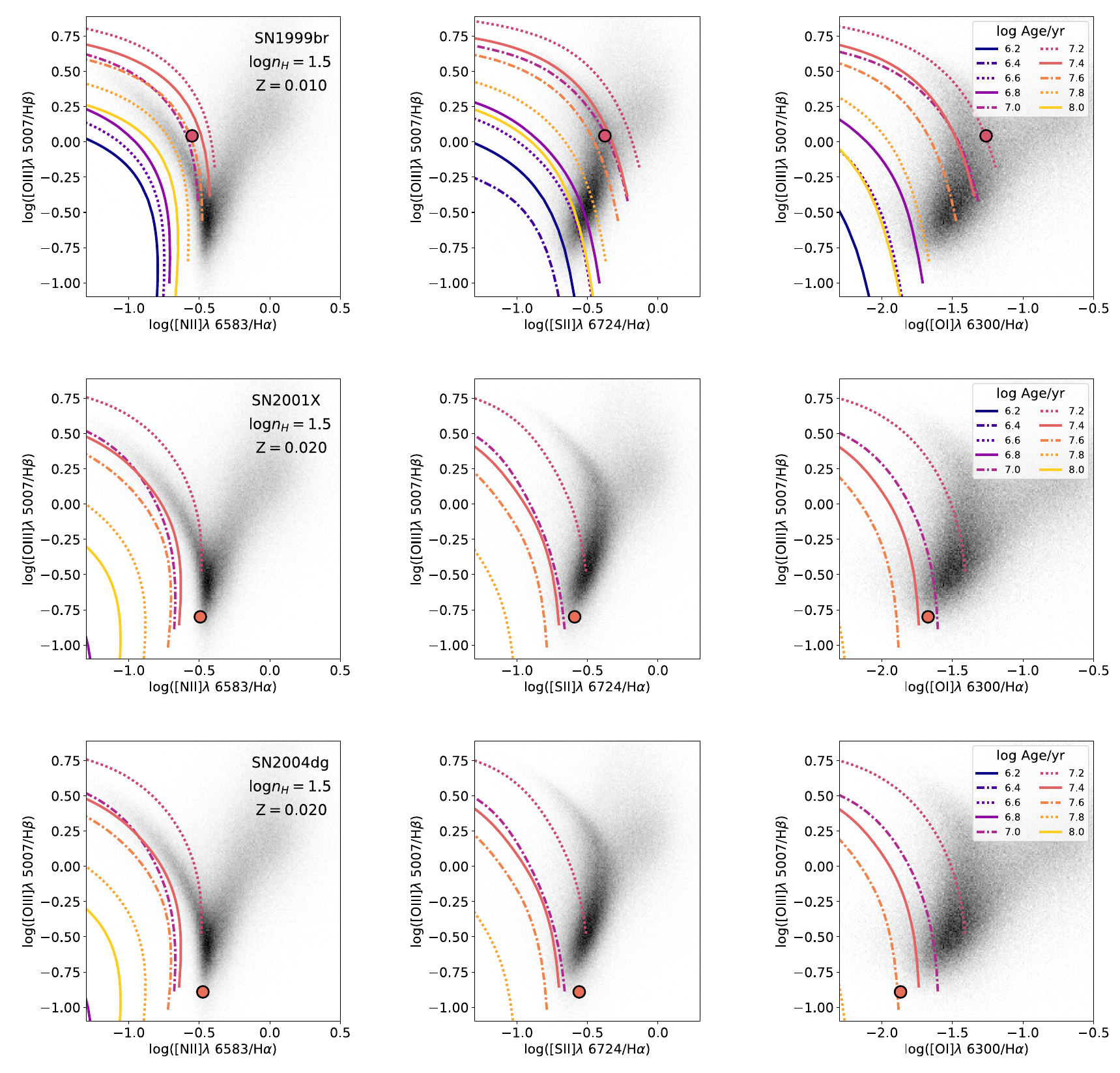}
\caption{BPT diagrams with [\ion{O}{iii}]/\hb\ against [\ion{N}{ii}]/\ha\ (left), [\ion{S}{ii}]/\ha\ (middle) and [\ion{O}{i}]/\ha\ (right) for each SN in our sample, one row per SN. In all panels, the distribution of SDSS galaxies is shown in grey scale, and the dashed curves correspond to BPASS binary system SPS models, which are colour coded according to stellar population age, indicated in the legend of the right-most panel. In each row, the metallicity and hydrogen density corresponding to the shown model curves is labelled in the first panel, and refers to the metallicity and n$_H$ that best matched the properties of the ionised gas at the SN position (filled circle). The SN data points are colour coded using the same key as the model curves, where the stellar progenitor ages correspond to the best-matched ages determined from the measured ionised gas line ratios and photon leakage-corrected \ha\ EW (see section~\ref{sssec:leakybinfits}). The SPS tracks trace the position of the models on the BPT diagram for decreasing ionisation parameters, $U$, starting from $\log(U)=-1.5$ (upper left) down to $\log(U)=-3.5$ (lower right).}\label{fig:leakyBPTs}
\end{figure*}

\begin{figure*}
\setcounter{figure}{1}
  \includegraphics[width=1.0\linewidth]{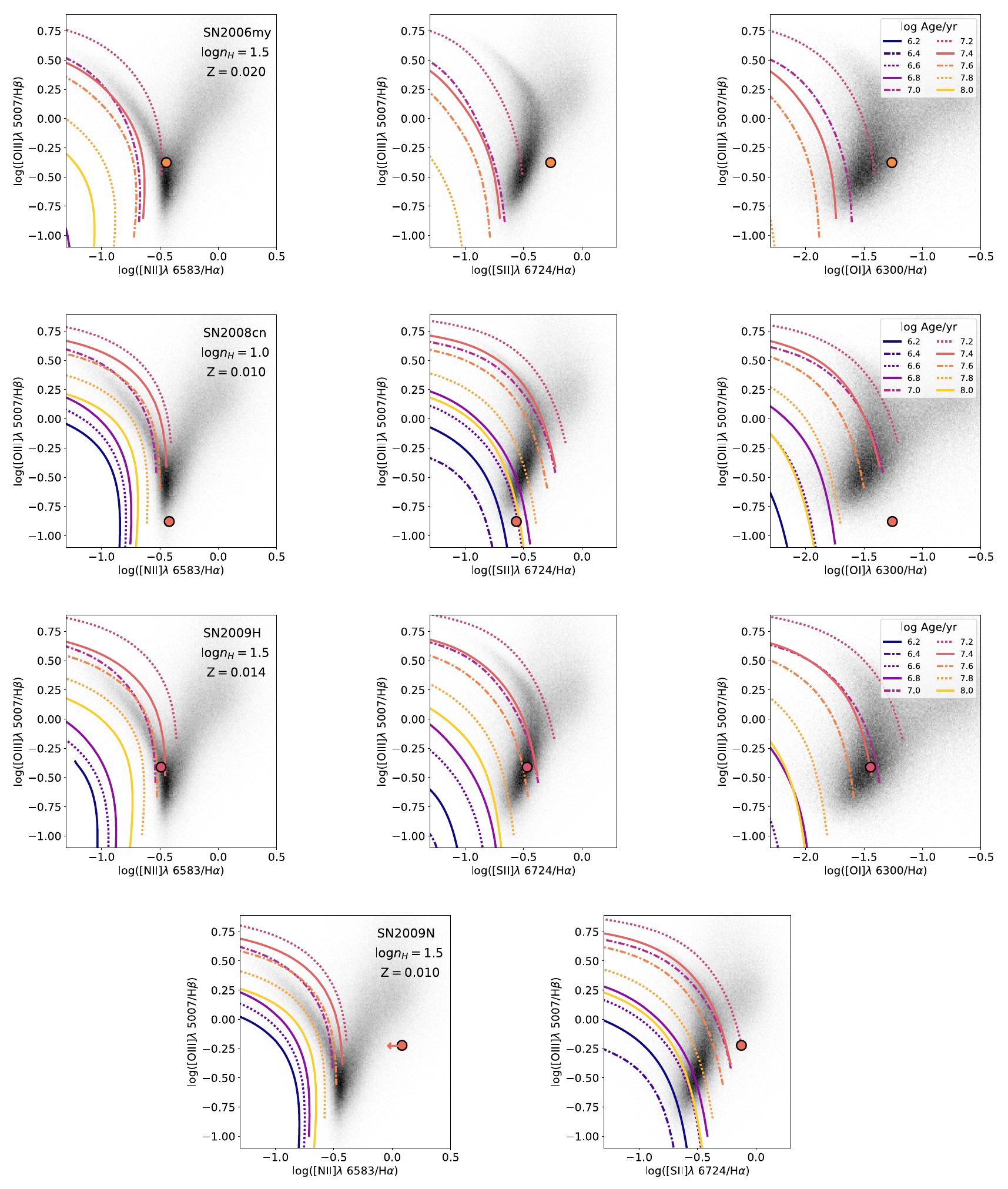}
\caption{cont. The [\ion{O}{i}]$\lambda$6300/\ha\ BPT diagram is not shown for SN~2009N because [\ion{O}{i}] is not detected at the SN position.}
\end{figure*}

\begin{figure*}
\setcounter{figure}{1}
  \includegraphics[width=1.0\linewidth]{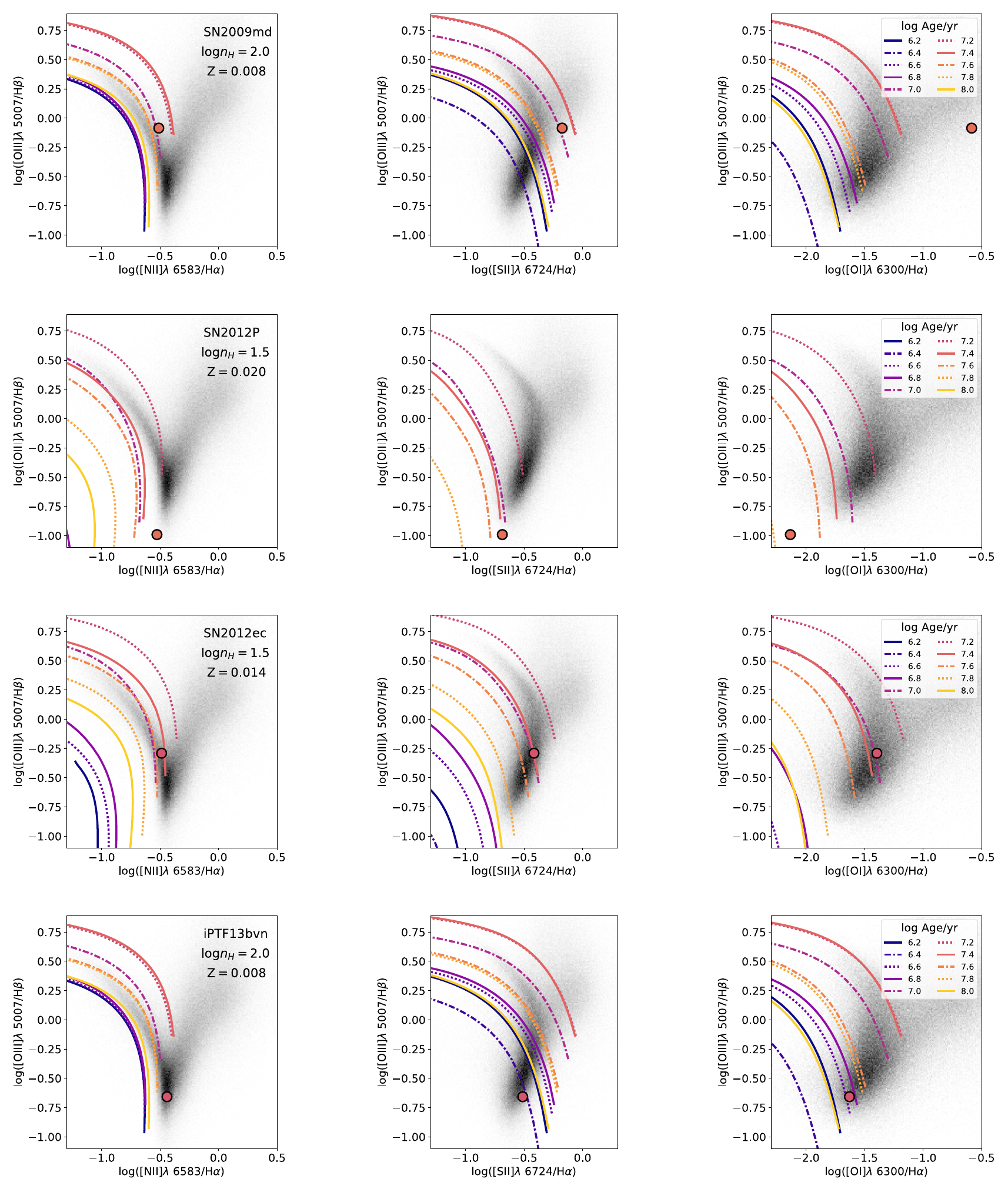}
\caption{cont.}
\end{figure*}

\bsp	
\label{lastpage}
\end{document}